\documentclass{aa}
\usepackage[varg]{txfonts}

\usepackage{natbib}
\bibpunct{(}{)}{;}{a}{}{,} 

\usepackage{graphicx}
\usepackage{hyperref}
\usepackage{bm}
\usepackage{placeins}
\usepackage{xcolor, color, colortbl}

\usepackage{amsmath}    
\usepackage{cuted}  
\usepackage{cleveref}
\usepackage{physics}

\defcitealias{Finlay2023}{Paper I}

\begin{document}

\title{TABASCAL: Removing multi-satellite interference from radio interferometry observations}


\titlerunning{TABASCAL}

\author{
    C.~Finlay\inst{\ref{inst:unige}} \and
    B.~A.~Bassett\inst{\ref{inst:uj},\ref{inst:wits},\ref{inst:uct}} \and
    M.~Kunz\inst{\ref{inst:unige}} \and
    N.~Oozeer\inst{\ref{inst:sarao},\ref{inst:rhodes}}
    }

\authorrunning{C.~Finlay et al.}

\institute{D\'epartement de Physique Th\'eorique and Center for Astroparticle Physics, Universit\'e de Gen\`eve, 24 quai Ernest Ansermet, 1211 Gen\`eve 4, Switzerland\\
\email{christopher.finlay@unige.ch}\label{inst:unige}
\and
Institute for Machine Intelligence and Neural Discovery (MIND), University of the Witwatersrand, Johannesburg, South Africa\label{inst:uj}
\and
School of Computer Science and Applied Mathematics, University of the Witwatersrand, Johannesburg, South Africa\label{inst:wits}
\and
Department of Mathematics and Applied Mathematics, University of Cape Town, Rondebosch, Cape Town, 7700, South Africa\label{inst:uct}
\and
South African Radio Astronomy Observatory, Liesbeek House Building, River Park, Gloucester Road, Mowbray, Cape Town 7700, South Africa\label{inst:sarao}
\and
Centre for Radio Astronomy Techniques and Technologies, Department of Physics and Electronics, Rhodes University, P.O. Box 94, Makhanda 6140, South Africa\label{inst:rhodes}
}

\date{Received date /
Accepted date }

\abstract{
In the first trajectory-based radio frequency interference (RFI) subtraction and calibration (TABASCAL) paper, we showed how to calibrate radio interferometers in the presence of RFI sources by simultaneously isolating the trajectories and signals of the RFI sources. In this paper, we show that we can accurately remove RFI (i.e. recover the astronomical signal) from simulated MeerKAT radio interferometry target data. We  are able to do so for a single frequency channel, corrupted by up to nine simultaneous satellites, with average RFI amplitudes varying from weak to very strong ($1-10^3$ Jy). Additionally, TABASCAL also manages to leverage the signal-to-noise ratio (S/N) of the RFI to phase-calibrate the astronomical signal. TABASCAL, effectively performs a suitably phased up fringe filter for each RFI source, which essentially allows for an   ideal removal of RFI across all RFI strengths. As a result, TABASCAL is able to reach image noises equivalent to the uncorrupted, no-RFI, case. For larger RFI amplitudes, the resulting image noise is 10x - 100x smaller than those from traditional RFI flagging methods such as AOFLAGGER. As a specific application, we show that point-source science with TABASCAL almost matches the no-RFI case with near perfect completeness for all RFI amplitudes. In contrast, the completeness of AOFLAGGER and idealised $3\sigma$ flagging drops below 40\% for strong RFI amplitudes, where recovered flux errors are approximately 10x - 100x worse than those from TABASCAL. Finally, we note that TABASCAL works for astronomical sources with both static and varying fluxes. 
}


\maketitle

\section{Introduction}\label{sec:intro}

Radio astronomy has profoundly transformed our understanding of the universe by enabling the study of celestial objects through the detection of radio waves. These extremely faint cosmic signals face a growing threat from radio frequency interference (RFI), which is becoming increasingly problematic as the radio spectrum becomes more crowded due to advancements in modern technology. Defined as unwanted radio emissions that disrupt astronomical observations \citep{Kocz2010}, RFI arises from various sources such as terrestrial broadcast systems, cellular networks, satellite communications, and everyday electronic devices. Thus, it presents a formidable challenge in radio astronomy, compromising the quality and quantity of astronomical data, and thereby hindering our exploration of the universe. As the sensitivity of modern radio telescopes continue to improve, these instruments face an increasingly invasive RFI environment. These radio signals, which often overlap with the frequency bands used for science, tend to obscure, distort, or mimic the faint astrophysical emissions astronomers aim to detect.

The proliferation of communication technologies has exacerbated this issue. For instance, satellite constellations designed for global internet coverage, such as SpaceX’s Starlink, have been shown to unintentionally emit RFI at low frequencies including in  the protected bands of radio astronomy 
\citep{DiVruno2023, Grigg2023, Bassa2024}. No doubt, the increasing prevalence of these satellite constellations will lead to a worsening of the RFI environment for telescopes across the globe. Similarly, terrestrial sources such as mobile phone networks and radar systems add to the pervasive RFI environment, particularly in densely populated areas. Although international guidelines, such as those established by the International Telecommunication Union, provide some safeguards for radio astronomy, the dynamic and evolving nature of RFI demands more adaptive and sophisticated mitigation strategies. 

This paper is structured as follows. Section \ref{sec:rfi_mit} discusses the current state of post-correlation RFI mitigation approaches. Section \ref{sec:tabascal} briefly describes the trajectory-based RFI subtraction and calibration (TABASCAL) algorithm and its intrinsic workings. In Section \ref{sec:concepts}, we describe the basics of Bayesian inference (i.e. Gaussian processes), how they are used   to interpolate, and how they can be used as a fringe rate filter. In Section \ref{sec:forward_model}, we describe the full Bayesian forward model, how we set our priors based on sound theoretical footing, some of the intricacies when modelling RFI sources accurately, and the computational considerations to allow this method to effectively scale up. In Section \ref{sec:post_approx}, we discuss parameter transformations used to effectively perform optimisation in such a large parameter space as well as an optional method to obtain posterior uncertainty estimates in a scalable manner. In Sections \ref{sec:sim_desc} and \ref{sec:eval_desc}, we describe the simulation set we generated to test TABASCAL and how we compare our results. In Section \ref{sec:vis}, we analyse the performance of TABASCAL in the raw, recovered visibilities. In Sections \ref{sec:imaging} and \ref{sec:pnt_src}, we discuss the performance of TABASCAL recovered visibilities in the image domain, both in terms of image noise and artefacts and point source recovery. TABASCAL's phase calibration capabilities are analysed in Section \ref{sec:phase_cal}. The robustness of the prior distribution and potential improvements of the method are discussed in Section \ref{sec:discussion}. Finally, in Section \ref{sec:conclusions}, we summarise the method and our key results. 

\section{RFI mitigation}\label{sec:rfi_mit}

There are numerous RFI mitigation strategies and in general it needs to be a multi-faceted approach at any given observatory. For reviews on RFI mitigation strategies see \citet{Kesteven2010}, \citet{Briggs2005} \& \citet{Fridman2001}. Among the various mitigation strategies we are interested in offline post-correlation strategies. Post-correlation approaches to RFI mitigation fall broadly into two categories: (1) flagging and (2) subtraction or removal. Flagging is the process of identifying contaminated data samples and flagging them in a way that they are not included in any further downstream data analysis. The second category is removal or subtraction, which our method TABASCAL falls into. 

Currently, flagging methods fall into three categories: (1) traditional, (2) machine learning, and (3) modified likelihood. One of the most prevalent methods for traditional flagging is AOFLAGGER \citep{offringa2010aoflagger}, based on the SUMTHRESHOLD \citep{Offringa2010} algorithm. In the past several years, there has been a lot of work on using machine learning for RFI flagging, such as the work of \citet{Mosiane2016} and, specifically, deep learning triggered by \citet{Akeret2017}. Notable deep learning examples are from \citet{Kerrigan2019} and \citep{VafaeiSadr2020} as they apply to radio interferometric data. Finally, standing as potential flagging counterparts to our method are modified likelihood models such as \citet{Mhiri2022,Leeney2023,Anstey2024,Mhiri2024}. We note that the BEAMS formalism, introduced in \cite{Kunz2007} to deal with supernova contamination, can enable the efficient computation of the full Bayesian posterior for RFI flagging. We refer to these methods as modified likelihood as they do not assume the standard Gaussian likelihood appropriate for thermal noise alone; rather, they  modify it to include a contribution from outliers such as RFI contamination.

RFI flagging is a staple at radio observatories across the world, however, flagging comes at a cost. For example, with respect to observations by the MeerKAT telescope \citep{Jonas2018},  \citet{Sihlangu2021} presented a detailed analysis of 200 TB of data, (around 1500 observation hours). This analysis showed that more than 23\% of all L-band data is currently lost to RFI, with 37\% of the band subject to persistent RFI, mainly from satellites. This particularly affects neutral hydrogen intensity mapping efforts in our local universe \citep{Cunnington2022, Engelbrecht2024}. From existing RFI mitigation approaches at the VLA telescope \citep{Napier1983} in the 1-5 GHz band, a maximum loss of the order of 30\% is expected \citep{selina2020rfi}. These losses impact the sensitivity required by astronomers to carry out their sciences and, in turn, require additional observing time to compensate for the contaminated data, thus reducing the overall efficiency of radio telescope usage. Data loss can significantly impact the calibration process in radio astronomy, potentially introducing errors in flux density measurements and source localisation \citep{Rau2009}. Epoch of reionisation (EoR) science is an area where precise reconstruction of the cosmic signal is of paramount importance as the signal is buried in the noise. \citet{Offringa2019} shows that even flags from real data applied to a RFI-free simulated dataset can introduce biases that inhibit our ability to detect the EoR. 

The alternative to flagging is subtraction. Over the years, many methods for RFI subtraction have been proposed. Some online methods include spatial nulling \citep{Kocz2010} and cancellation using a reference horn \citep{Mitchell2005}. In the offline setting, there is subspace projection \citep{Shiyu2016}, post-correlation filtering \citep{Offringa2012, Helmboldt2019}, fringe fitting \citep{Athreya2009}, fringe rate CLEANing \citep{Kogan2010}, and deep learning \citep{Zhang2024}. Spatial nulling and filtering depends on the instrument having a multi-beam receiver. Although effective, this limits its use to such instruments, however, it comes at the cost of creating primary beam irregularities. Subspace projection methods also show great promise; however, they no longer show good separation performance when the RFI signal becomes decorrelated within a single time dump. This limits their applicability as many RFI signals decorrelate on the typical integration time scales used. Reducing these times increases data usage dramatically and can quickly become infeasible. Post-correlation filtering, fringe-fitting, and fringe rate CLEANing all depend on the difference in fringe rate between the signal-of-interest and the RFI signal. These are powerful techniques, however, they operate on a single baseline. Since the differential fringe rate can be low on certain baselines, these methods become ineffective in this case. In many instances, this is also where the RFI strength is highest. A method is thus required that avoids many of these pitfalls. 

Our proposed method, TABASCAL, is an extension of the method by the same name introduced in \citet[][hereafter Paper I]{Finlay2023}. TABASCAL performs an antenna decomposition of the signal from RFI sources, thus taking advantage of all baseline information simultaneously to separate the astronomical and RFI signals. TABASCAL is in essence a progression, although extensive, of the method proposed by \citet{perley2003removing} with additional features such as the potential to calibrate off the RFI signal. Additionally, TABASCAL avoids the need for high time resolution data and therefore is applicable to more standard post-correlation data, including archived observations. 

TABASCAL draws on ideas from the work of \citet{Roth2023} with respect to Gaussian processes, but places them in the context of fringe filtering and fitting. It is conceivable to envision a single method combining TABASCAL with the work of \citet{Roth2023} (imaging) and \citet{Leeney2023} (flagging unmodelled effects) to form a unified Bayesian approach to the radio interferometric data reduction process.

\section{Overview of TABASCAL }\label{sec:tabascal}

TABASCAL is a method used to separate RFI from astronomical signals in post-correlation radio interferometry data in an offline setting. Additionally, under the right circumstances, it can also jointly perform phase calibration. It employs a Bayesian framework to effectively cast the use of fringe rate filters as Bayesian priors. In this way, multiple filters can be applied simultaneously in multiple directions.

\subsection{TABASCAL I and II}\label{sec:IvsII}

This paper is the second in a series. TABASCAL I (hereafter \citetalias{Finlay2023})  addressed the problem of deriving antenna gains from calibrator observations in the presence of RFI moving on regular trajectories relative to the phase centre (e.g. satellites, towns, etc.). Therefore, TABASCAL I operates in the situation where the astronomical visibilities are known (i.e. a calibrator source) and the unknown quantities to be estimated are the antenna gains and the RFI signal parameters. In TABASCAL II, the astronomical visibilities are also unknown. This is the primary distinction between TABASCAL I and II. Currently, TABASCAL estimates a single polarisation product, for example, the XX polarisation, across all antenna pairs (baselines). In the future, TABASCAL will take advantage of the full polarisation nature of the data to enhance its separation capabilities as RFI is often strongly polarised compared to astronomical signals.

Traditionally, the antenna gains, derived from the calibrator observation, would be applied to a target observation in a transfer calibration sense; namely, first-generation calibration (1GC) as defined in \citet{Smirnov2011b}. In contrast, in this paper, titled TABASCAL II, we are analysing the target science observations directly. Calibration using the target observation data directly is referred to as self-calibration (selfcal) or second-generation calibration (2GC) in \citet{Smirnov2011a}. In 2GC, the unknown quantities are the antenna gains as well as the astronomical visibilities. In our case, we also have unknown RFI in the target observations.  Traditional self-calibration is an iterative, cyclic process. Initially, gains are fixed and the visibilities are imaged. The imaging process acts like a prior in the sense that it searches for solutions that are sparse in the image domain. After this, astronomical visibilities are fixed for each round when calculating gain solutions such that the problem is over-constrained \footnote{During selfcal, the problem is only over-constrained when there are more than three antennas. This is almost always the case.}. In contrast, TABASCAL II solves for the astronomical visibilities, antenna gains and the RFI signals jointly in one pass. As such, TABASCAL II requires prior information. The prior information provided for the astronomical and RFI signals is in the form of the scale of the signal and its time variability, namely, fringe rate filters. 

As described above, TABASCAL I and II solve a similar problem but in two different situations, namely, calibrator and target observations. They employ largely the same data model, however, there are some notable differences in TABASCAL II\ versus TABASCAL I, respectively:

\begin{enumerate}
    \item Astronomical visibilities are modelled versus known a priori;
    \item Estimated RFI signals are complex-valued versus real-valued;
    \item Satellite trajectories are calculated using accurate two-line elements (TLEs) instead of the simpler circular orbit model;
    \item RFI trajectory errors are absorbed into the complex-valued RFI signal versus a parametrised trajectory model;
    \item RFI signal interpolation is performed using a time-correlated covariance function versus linear interpolation;    \item Antenna gains are interpolated to observation times versus a parameter for each time step;
    \item We did not fit for the trajectories of the satellites here, assuming them to be previously known within some degree of error. If necessary, discovery can be performed using the methods of the TABASCAL I paper.  
\end{enumerate}

All of the differences noted above (except the first, i) are improvements on the previous work and can be retroactively applied there.

\subsection{TABASCAL II's working principle}\label{sec:how}

To learn the RFI signal, gains and astronomical visibilities, TABASCAL II effectively applies a fringe rate filter in multiple directions simultaneously, jointly fitting for the signals from these different directions. These fringe rate filters are applied through the use of the appropriate, derived priors on the associated parameters. By design, the astronomical visibilities for a tracking, fringe stopped interferometer, with a limited field of view, are expected to have a fringe frequency very close to zero \citep{Offringa2012}. The maximum expected fringe rate is proportional to the projected baseline length and the field of view of the telescope. For RFI sources, we manually fringe stop in the direction of each source and then apply a fringe rate filter in this direction. Fringe stopping in a particular direction causes the fringe rate for that source to be close to zero. To effectively separate RFI sources whose fringe rate is close to that of the astronomical sources or each other, each RFI source has its signal modelled at antenna level. When doing this, we need to account for fringe winding loss as RFI signals will decorrelate when the integration time is sufficiently long compared to the source's fringe rate. We do this by interpolating the RFI signals to sub-integration time scales and then integrating back to the data rate. Using an antenna based decomposition allows us to account for direction dependent gains in the direction of that specific source that may vary across antennas, such as the primary beam, ionospheric effects, and position errors. In essence, TABASCAL extends the concept of peeling \citep{Intema2009} to RFI sources. The signal separation capabilities of TABASCAL are hinged on fringe rate filtering and peeling of a time-varying source. 

TABASCAL II\ has phase calibration capabilities that depend on the signal-to-noise ratio (S/N) of the RFI signal. This is due to the sensitivity of the predicted visibilities to small changes in the position of the RFI sources. This is thanks to the antenna decomposition of the signal for these sources. The priors used on the astronomical and RFI signal strengths are non-informative. As such, TABASCAL II\ does not perform amplitude calibration. However, this is possible if a known source is in the field or more a priori information is known about an RFI source. As for phase calibration, TABASCAL II\ 's capabilities are dependent on the strength of the RFI sources. If the signal is too weak, we do not have a  high-enough S/N. In this case, TABASCAL II\ should return gain estimates that are consistent with the prior which is set by traditional 1GC or the use of TABASCAL I when RFI sources are present.

\section{Method}\label{sec:method}

Our proposed method, TABASCAL, makes use of a Bayesian forward model to predict the RFI contaminated visibilities from a set of parameters (including nuisance parameters) to model the RFI signal. The parameters of this model are then estimated by optimising over the posterior distribution to obtain a maximum a posteriori (MAP) estimate. Thereafter, Gaussian constrained realisations can be used to estimate the posterior covariance. All nuisance parameters can be ignored, leaving us with calibrated astronomical visibilities along with optional error estimates. All time-varying signals, such as antenna gains, astronomical visibilities, and the RFI signal at the antennas, are modelled using Gaussian processes where prior knowledge of their time variability can be encoded. We expect the antenna gains to vary on a longer time scale compared to the astronomical visibilities which in turn are expected to vary more slowly than the RFI signal. Encoding this prior information into the probabilistic model allows us to break the inherent degeneracies when estimating more parameters than data points, as is the case here. 

In this section, we  review the required concepts to build our Bayesian forward model and how to obtain a MAP and covariance estimate. We start, in Section \ref{sec:concepts}, by introducing Bayes theorem, Gaussian processes (GP), and two different implementations of GPs. Next, in Section \ref{sec:forward_model}, we introduce our full Bayesian forward model with all the separate components including their prior distributions. We also explain how to determine the prior distributions and finish by discussing the computational considerations for the two different GP methods. Finally, in Section \ref{sec:post_approx}, we explain how optimisation is performed using standardised parameters as well as how to estimate the posterior covariance in a scalable manner.

\subsection{Concepts}\label{sec:concepts}

\subsubsection{Bayesian overview}\label{sec:bayes}

The central object in Bayesian statistical methods is the posterior distribution, $\mathcal{P}(\bm{\theta}|\mathcal{D})$, namely, the probability distribution over model parameters, $\bm{\theta}$, given some observed data, $\mathcal{D}$. Bayes theorem gives
\begin{equation}\label{eq:bayes}
    \mathcal{P}(\bm{\theta}|\mathcal{D}) = \frac{ \mathcal{L}(\mathcal{D} | \bm{\theta}) \Pi(\bm{\theta}) }{ \mathcal{Z}(\mathcal{D}) },
\end{equation}
which is constructed from the likelihood, $\mathcal{L}(\mathcal{D} | \bm{\theta})$, the prior, $\Pi(\bm{\theta})$, encoding our beliefs about the parameters before seeing the data, and the evidence or marginal likelihood, $\mathcal{Z}(\mathcal{D})$, which is a normalising constant that is not required to find the MAP estimate or to draw samples from the posterior. The likelihood term is derived from the noise distribution that we assume in our data model and is what links information from our data to our model parameters, $\bm{\theta}$. The prior distribution is where we can include any additional information to drive our solutions in a desired direction from information we have about the problem without the knowledge of the data. The fundamental piece of prior information we use in this analysis is the expected time variability of the different signals present in our observations.

\subsubsection{Gaussian process}\label{sec:gp}

A Gaussian process is a collection of random variables, any finite number of which have a joint multivariate normal distribution. It defines a distribution over functions, in the sense that for any finite set of input points, the corresponding outputs follow a joint Gaussian distribution. In our work, we consider GPs defined over a one-dimensional temporal domain. Analogous to the normal distribution, a GP is fully specified by its mean function and covariance function. The covariance function encodes the pairwise correlations between function values at different time points.

A common class of covariance functions is the Mat\'ern class of covariance functions. The squared exponential (SE) is the limiting case of these and leads to the smoothest functions. In this work, we use the SE covariance function which is defined as
\begin{equation}\label{eq:SE_cov}
    \kappa(t_1,t_2) = \sigma^2 \exp \left[ \frac{-|t_1-t_2|^2}{2\tau^2} \right].
\end{equation}

Therefore, prior knowledge of the time variability, between $t_1$ and $t_2$, and the overall scale of a function and/or signal can be encoded through the definition of an appropriate covariance function. Then, $\tau$ determines the length scale of the GP solutions and is referred to as a hyperparameter in the context of our work. 

For example, in this work we wish to estimate the gain phase values over the observation period. If we expect these to vary by less than $10^\circ$ per hour \footnote{Example gain solutions for MeerKAT are shown on \href{https://ragavi.readthedocs.io/en/latest/_images/gplot.png}{https://ragavi.readthedocs.io/en/latest/\_images/gplot.png}}, then we would choose $\sigma=10^\circ$ and $\tau_{\phi_G}=1$ hour as a prior covariance between the gain phase values on a single antenna. In this work these parameters are determined by a fit to data from adjacent calibration observations which is covered in \citetalias{Finlay2023}. Alternatively, these parameters could also be varied and have an associated hyperprior on them based on the estimate from the calibration observations.

\subsubsection{Interpolation and inducing points}\label{sec:interpolation}

GP regression can be viewed as an interpolation when the covariance of the known points is assumed to be noise-free. Given a covariance function, $\kappa(t_1,t_2)$, locations of known points, $\bm{t}'$, and new, interpolated function locations, $\bm{t}$, an interpolator, $\bm{I}_{tt'}$, can be defined as
\begin{align}\label{eq:interpolator}
\begin{split}
    \bm{y}(\bm{t}) &= \bm{C}_{tt'} \bm{C}_{t't'}^{-1} \bm{y}' \\
    &= \bm{I}_{tt'} \bm{y}' .
\end{split}
\end{align}
The evaluation of the covariance function, $\kappa(t_1,t_2)$, over all combinations of locations in $\bm{t}'$ gives the covariance matrix, $\bm{C}_{t't'}$. Then, $\bm{y}$ represents the interpolated function values at the points, $\bm{t}$, and $\bm{y}'$ represents the function values at the locations, $\bm{t}'$. Furthermore, $\bm{C}_{t't'}$ is the covariance of the prior distribution of the function values at the locations, $\bm{t}'$. In this work, the values $\bm{y}'$ and locations $\bm{t}'$ are referred to as inducing points \citep{lawrence2002fast}. The $\bm{y}'$ values will be parameters in our model and subsequently estimated. The $\bm{t}'$ locations are fixed in our model and regularly spaced at intervals of approximately $\tau$ with the endpoints of our observation interval included. The $\bm{y}$ values are the interpolated points which could be at the data rate as is the case for our gains, or at a higher sampling rate as is the case for our RFI signals.

\subsubsection{Fourier-based GP method}\label{sec:fourier_gp}

A particular class of covariance functions commonly used are homogeneous/stationary covariance functions, namely, they depend only on the difference in time. Mat\'ern covariance functions form part of this class. Evaluating a homogeneous covariance function on a periodic domain, with regular spacing, leads to a circulant covariance matrix. Such a matrix is trivially diagonalisable with a Discrete Fourier Transform which can be implemented using the fast Fourier transform (FFT). The Wiener-Khinchin theorem states that the Fourier transform of a stationary covariance function is the power spectrum of the Gaussian process (GP). The relation between the covariance matrix, $\bm{C}_{tt}$, and the power spectrum, $\bm{P}_{\eta\eta}$, is given by
\begin{equation}\label{eq:fourier_cov}
    \bm{C}_{tt} = \bm{F} \bm{P}_{\eta\eta} \bm{F}^H,
\end{equation}
where $\bm{F}$ is the Fourier transform, $\bm{F}^H$ is its Hermitian transpose (inverse), and $\bm{P}_{\eta\eta}$ is a diagonal matrix. Then, $\bm{P}_{\eta\eta}$ is therefore the covariance in Fourier space $\eta$, namely, the fringe rate space when used for visibilities as is the case for us.

\subsection{Bayesian forward model}\label{sec:forward_model}

To compute the likelihood in Equation \eqref{eq:bayes}, we need a model that predicts the data given model parameters. In this work we simulate data from a radio interferometer, referred to as visibilities. Visibilities are a measure of the signal coherence from the sky in two locations, namely, the locations of a pair of antennas called a baseline. The visibility for an ideal interferometer is defined as
\begin{equation}\label{eq:ast_vis}
    V \left( \vec{u}, \lambda \right) = \iint_{lm} I \left( \vec{l} \right) \exp \left[ -\frac{2\pi i}{\lambda} \left( \vec{l}-\vec{l_0} \right) \cdot \vec{u}  \right] \frac{dldm}{n},
\end{equation}
where $V$ is the visibility, I is the sky surface brightness distribution, $\vec{l}=(l,m,n)$ are the sky coordinates, $\vec{u}=(u,v,w)$ are the baseline coordinates, and $\lambda$ is the observation wavelength. Here, $\vec{l}_0$ is the direction of phase tracking centre and is used to fringe stop in that direction. The visibilities can be thought of as the Fourier transform of the sky surface brightness. This becomes true in the limiting case when the telescope is co-planar $(w=0)$ or we only consider a small field of view $(n \approx 1)$. In this context, we are only dealing with the visibilities in our proposed method; however, the link to the sky should not be forgotten and will be used in our analysis of the method. 

Our visibility model for the RFI contaminated visibilities takes the simple functional form of
\begin{equation}\label{eq:vis_model}
    V_{pq} = G_p \left( V_{pq}^\text{AST} + V_{pq}^\text{RFI} \right) G^*_q ,
\end{equation}
where $G_p$ are the complex-valued antenna gains on the antenna, $p$, while $V_{pq}^\text{AST}$ represents the astronomical visibilities and $V_{pq}^\text{RFI}$ are the RFI visibilities on the baseline formed by the antennas $p$ and $q$. All of these terms vary over time but at different rates. The complex-valued noise in the visibilities is assumed to be independent and identically distributed Gaussian noise with standard deviation of $\sigma_n$ in both the real and imaginary components \citep[Chapter~6]{Thompson2017}. The likelihood is therefore 
\begin{equation}\label{eq:likelihood}
    p(\bm{V}^\text{OBS} | \bm{\theta}) = \left( 2 \pi \sigma_n^2 \right)^{-N_D} \exp \left[ \frac{ | \bm{V}^\text{OBS} - \bm{V}\left( \bm{\theta} \right) |^2}{ 2\sigma_n^2} \right] ,
\end{equation}
where $\bm{V}^\text{OBS}$ is the vector of all observed visibility data, $\bm{V}$ is the vector of model visibilities, $\bm{\theta}$ is the vector of all the model parameters, and $N_D$ is the number of data points (i.e. the dimensionality of $\bm{V}^\text{OBS}$). Therefore, $N_D = N_T N_A (N_A-1) / 2$. $N_T$ is the number of time steps in the observation and $N_A$ is the number of antennas in the array. It should be noted that the likelihood in Equation \eqref{eq:likelihood} is for complex values and, therefore, $N_D$ is the number of complex-valued data points. Finally, $\sigma_n$ is the noise on the real and imaginary components individually. 

\subsubsection{Astronomical visibility model}\label{sec:ast_vis_model}

The astronomical visibilities are modelled with a GP in Fourier (fringe rate) space, $\tilde{V}_{pq}(\bm{\eta})$. Then an inverse Fourier transform is applied to obtain functions of time, $V_{pq}(\bm{t})$, for each individual baseline. The power spectrum therefore defines the prior covariance in Fourier space, and the prior mean for each component is 0, namely,
\begin{equation}\label{eq:vis_ast_prior}
    p \left( \tilde{V}_{pq}^\text{AST} \right) = \mathcal{N} \left( \bm{0}, \bm{P}_{pq} \right),
\end{equation}
where $\mathcal{N}(\bm{\mu}, \bm{\Sigma})$ represents the multivariate normal distribution with mean $\bm{\mu}$ and covariance $\bm{\Sigma}$. $\bm{P}_{pq}$ is diagonal and is the expected power spectrum of $V_{pq}(\bm{t})$. Obtaining the astronomical visibilities over time is done with
\begin{equation}\label{eq:vis_ast_transform}
    V_{pq}^\text{AST}(\bm{t}) = \bm{F}^H \tilde{V}_{pq}^\text{AST}(\bm{\eta}).
\end{equation}

\begin{figure}
  \centering
  \includegraphics[width=0.5\textwidth]{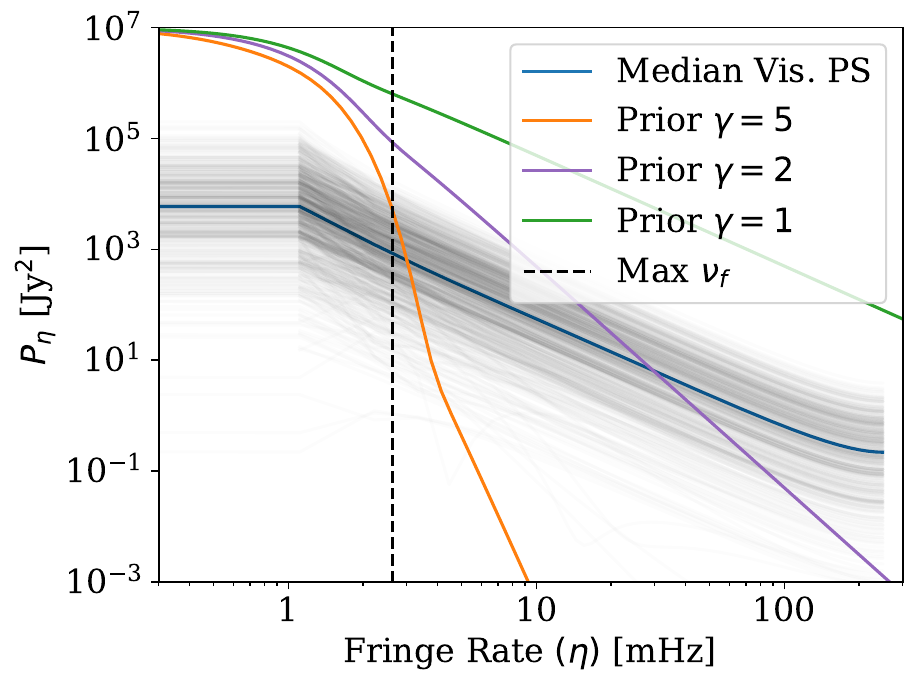}
  \caption{Power spectrum prior used for our astronomical visibilities (orange). The faint black curves are the true calculated power spectrum for all baselines in one of our simulations. The blue curve shows the median power spectrum and the black dashed line shows the maximum expected fringe rate of the longest baseline in our simulation. The parameters for the example priors shown are $P_0=10^7$ Jy$^2$, $\eta_0=1$ mHz, and $\gamma=5$, 2, and 1.}
  \label{fig:ps_ex}
\end{figure}

The power spectrum is the squared magnitudes of the visibilities in fringe rate space. The analytical form of the power spectrum used in this work is
\begin{equation}\label{eq:power_spec}
    P(\eta) = \frac{P_0}{2} \left[ \exp(-\frac{\eta^2}{2\eta_0^2}) + \left( 1 + \frac{\eta^2}{\eta_0^2} \right)^{-\gamma} \right],
\end{equation}
where $P_0$ controls the overall power of the signal, namely, the variance of the signal in Fourier space. This power spectrum essentially looks like a leg with a bent knee. Here, $\eta_0$ controls the position of the knee in $\eta$-space and $\gamma$ controls the angle of the lower leg. Since the angle of the lower leg determines the proportion of lower and higher frequency modes in the signal, $\gamma$ therefore controls the smoothness of the signal with larger values increasing smoothness. Figure \ref{fig:ps_ex} (orange, purple, and green curves) shows examples of the power spectrum. Then, $P_0$ should be chosen based on our expectation of the sky signal itself. We  used $P_0=10^7$ throughout this work, which corresponds to a prior standard deviation of $\sqrt{P_0}/N_T = 7$ Jy. Here, $\gamma$ is user-tunable and exists mostly for numerical stability, however, it should remain above 2. Due to our knowledge of the instrument and our pointing direction, we can appropriately choose the $\eta_0$ parameter as simply the maximum expected fringe rate for a given baseline. Excluding any strong astronomical sources in our sidelobes, the maximum fringe rate for a baseline would be expected when a strong source is on the edge of our field of view (FoV). In this case, the fringe rate is given by
\begin{equation}\label{eq:ast_fringe}
    \nu_f = \omega_E \frac{|\vec{u}|}{\lambda} \sin \left( \theta_\text{FoV} / 2 \right),
\end{equation}
where $\omega_E$ is the rotation rate of the Earth in rad.s$^{-1}$, $|\vec{u}|$ is the baseline length, $\lambda$ is the observation wavelength and $\theta_\text{FoV} = 1.22 D/\lambda$ is the field of view of the telescope with a dish diameter of $D$. For our simulations with a maximum baseline of 800 m, 25 cm wavelength, and a 13.5 m dish diameter, this works out to $\nu_f = 2.6$ mHz. We have used $\eta_0=1$ mHz throughout this work.

In practice, because the FFT forces periodic solutions on a finite interval, we end up with edge effects in the estimated astronomical visibilities in the time domain. Additionally, many baselines have fringe periods much larger than the simulated observations we are using. To remedy this, we make use of padding.

\subsubsection{Gain model}\label{sec:gain_model}

The antenna gains are modelled with a GP using inducing points $G_p(\bm{t}')$ which are then interpolated to the points $G_p(\bm{t})$. This is done using the interpolator defined by their covariance function as is described in Section \ref{sec:interpolation}. A separate interpolator is used for the gain amplitudes compared to the gain phases as these are expected to vary at different rates and have different variances. However, the same interpolator is used for all antennas as these are expected to vary at the same rate. Therefore, 
\begin{equation}\label{eq:gain_amp_interp}
    |G(\bm{t})| = \bm{I}^{|G|} |G(\bm{t}')|, 
\end{equation}
and
\begin{equation}\label{eq:gain_phase_interp}
    \phi_G(\bm{t}) = \bm{I}^{\phi} \phi_G(\bm{t}').
\end{equation}
The gain amplitudes $|G|$ and phases $\phi_G$ are then combined as 
\begin{equation}\label{eq:gains}
    G = |G|\exp \left[ i \phi_G \right] 
\end{equation}
to form the complex gains at each antenna over time.

The prior distribution for the antenna gains is chosen based on the estimates obtained from the expected calibration observations that would sandwich the target observation in question. A GP can be fitted to the calibration estimates at the times $\bm{t}''$ of the calibration observations in the standard way as is well described in \citet{Rasmussen2005}. We briefly summarise the procedure below. 

Given gain estimates $\bm{y}''$ with error covariance $\bm{\Sigma}''$ at calibration times $\bm{t}''$ we can fit the appropriate SE kernel parametrised by $l$ and $\sigma$, giving us $\bm{C}''=\bm{C}_{t''t''}$. The gain estimates from the calibration observation can be obtained, even in the presence of RFI, using the method described in \citetalias{Finlay2023}. The prior mean $\bm{\mu}'$ and prior covariance $\bm{\Sigma}'$ for our inducing points $\bm{y}'$ are therefore given by
\begin{equation}\label{eq:mean_update}
    \bm{\mu}' = \bm{C}_{t't''} \left( \bm{C}'' + \bm{\Sigma}'' \right)^{-1} \bm{y}'',
\end{equation}
and
\begin{equation}\label{eq:cov_update}
    \bm{\Sigma}' = \bm{C}' - \bm{C}_{t't''} \left( \bm{C}'' + \bm{\Sigma}'' \right)^{-1} \bm{C}_{t't''}^T.
\end{equation}
The prior distribution for our inducing points $\bm{y}'$ at the times $\bm{t}'$ is then 
\begin{equation}\label{eq:GP_prior}
    p(\bm{y}') = \mathcal{N} \left( \bm{\mu}', \bm{\Sigma}' \right).
\end{equation}
The priors for the gain amplitudes and gain phases are assumed independent of each other, as is the case for all prior terms that make up the full Bayesian model. When calculating the prior covariances and interpolators $\bm{I}^{|G|}$ and $\bm{I}^{\phi}$ via Equation \eqref{eq:interpolator}, the $\sigma$ and $l$ values estimated from fitting a GP to the calibration portions are used.

\subsubsection{RFI visibility model}\label{sec:rfi_vis}

The RFI visibility model is constructed from two parts, these are then combined and then averaged to effectively model fringe winding (time-smearing) of the RFI visibilities. The complex-valued RFI signal at each antenna is modelled as a GP using inducing points and the trajectories of the RFI satellite sources are modelled using two-line element sets (TLEs) with a simplified general perturbation (SGP) model \citep{Hoots1980}. From the trajectory, geometric phase delays are calculated. For a real observation, the actual position of the satellite can differ from the predicted position using TLEs. This position error leads to a differential fringe rate and phase offset between the model and the truth. When this error is not too large, this can be accounted for by the fitted complex signal at each antenna. Alternatively, an implementation of SGP in the auto-differentiation framework PyTorch has already been released by \citet{Acciarini2025} and could be included in the future to account for this.

The overall RFI visibility model looks like
\begin{equation}\label{eq:rfi_vis}
    V^\text{RFI}_{pq}(t) = \frac{1}{\Delta t} \int^{t+\Delta t /2}_{t-\Delta t /2} \sum_s A_{ps}(t) A_{qs}^*(t) K_{ps}(t) K_{qs}^*(t) dt, 
\end{equation}
where $\Delta t$ is the integration time of each observed data point, $A_{ps}(t)$ is the RFI signal at antenna $p$ from the source labelled $s$, and $K_{ps}(t)$ is the geometric phase delay term induced by the trajectory of the RFI source. Each RFI source is modelled as a point source, this is valid for nearly all RFI sources except those present within the telescope site \citepalias{Finlay2023}.   

The integral in Equation \eqref{eq:rfi_vis} is calculated using a left Riemann sum with the sampling points defined with regular spacing. The sampling frequency is determined by the component with the fastest time variability which is typically the geometric phase delay term on the longest baseline. The sampling rate, $\nu_s$, required to maintain closure relations \citet{perley2003removing} is 
\begin{equation}\label{eq:rfi_sampling}
    \nu_s > \pi \lvert \nu_f \rvert \sqrt{\frac{ \lvert V_\text{Inst}^\text{RFI} \rvert }{6 \sigma_n}},
\end{equation}
where $ \vert V_\text{Inst}^\text{RFI} \rvert$ is the maximum instantaneous visibility amplitude of a source and $\nu_f$ is its fringe frequency. In addition to this, there is an error, $\epsilon$, introduced by the numerical integration. For a Riemann sum, this is
\begin{equation}\label{eq:riemann}
    \epsilon \leq \frac{ \lvert V_\text{Inst}^\text{RFI} \rvert \Delta t}{\nu_s}.
\end{equation}
Thus, to maintain an integration error below the visibility noise, $\sigma_n$, we have an upper bound defined by the numerical integration scheme used. This results in bounds for $\nu_s$ as 
\begin{equation}\label{eq:rfi_bounds}
    \pi \lvert \nu_f \rvert \sqrt{\frac{ \lvert V_\text{Inst}^\text{RFI} \rvert }{6 \sigma_n}} < \nu_s \leq \frac{ \lvert V_\text{Inst}^\text{RFI} \rvert \Delta t}{\sigma_n}.
\end{equation}
This bound can be improved upon if using a more accurate integration scheme such as Simpson's rule; thus, leading to a more computationally efficient implementation. 

The fringe frequency is determined by the observation wavelength, $\lambda$, the baseline length, $B$, the velocity of the RFI source relative to the baseline, and the distance of the RFI source from the baseline. The fringe frequency of a moving RFI source is given by
\begin{equation}\label{eq:sat_fringe}
     \nu_f = -\frac{1}{\lambda} \left( \frac{\left( \vec{r}_s - \vec{r}_p \right) \cdot \dot{\vec{r}}_s}{\lvert \vec{r}_s-\vec{r}_p \rvert} - \frac{\left( \vec{r}_s - \vec{r}_q \right) \cdot \dot{\vec{r}}_s}{\lvert \vec{r}_s-\vec{r}_q \rvert} + \omega_E u_{pq} \cos \delta_0 \right),
\end{equation}
where $\vec{r}_s$ and $\dot{\vec{r}}_s$ are the position and velocity of the RFI source labelled $s$, respectively, and $\vec{r}_p$ and $\vec{r}_q$ are the positions of the antennas labelled $p$ and $q$ respectively. These should be expressed in the International Terrestrial Reference Frame coordinates. Then, $\omega_E$ is the rotation rate of the Earth, $u_{pq}$ is the $u$-component of the baseline, $pq$, and $\delta_0$ is the declination of the phase centre. The first two terms are due to the motion of the satellite and the last term is due to the natural fringe frequency induced by the rotation of the Earth \citep{Thompson2017}. For an RFI source that is stationary with respect to the telescope, $\dot{\vec{r}}_s=\vec{0}$, and therefore only the last term in Equation \eqref{eq:sat_fringe} remains. This is the more common result shown \cite[Chapter~4.3]{Thompson2017} and other works.

\subsubsection{RFI geometric phase delays}\label{sec:rfi_geo_phase}

The RFI geometric phase delay model is a per antenna term that is dependent on the trajectory of the RFI source. In this work, we consider RFI sources to be in the near-field. Given than the Fraunhofer distance for an interferometer is
\begin{equation}
    d_F=2|\bm{u}|^2/\lambda,
\end{equation}
where $|\bm{u}|$ is the baseline length and $\lambda$ is the observational wavelength. For the MeerKAT telescope with baselines up to around 8 km and maximum wavelength of approximately 1 m, the Fraunhofer distance is 128,000 km. This is well beyond the orbital height of even geostationary satellites. As such, to maintain the correct phases for the longest baselines we need to treat all RFI sources as near-field. The computational cost of using a near-field formulation is larger than that of a far-field but overall becomes a negligible cost in comparison to the calculation of per baseline visibilities from antenna-based signals for the RFI sources. This is due to it scaling with the square of the number of antennas. The geometric phase delay on a single antenna is defined as
\begin{equation}\label{eq:rfi_phase}
    K_{ps}(t) = \exp \left[ - \frac{2\pi i}{\lambda} \left( |\vec{r}_s(t) - \vec{r}_p(t)| + w_p(t) \right)  \right],
\end{equation}
where $\vec{r}_s(t)$ is the position of the RFI source labelled $s$, $\vec{r}_p(t)$ is the position of the antenna labelled $p$, and $w_p$ is the $w$-component of the antenna position. The $w$-component term comes from the delay compensation introduced in a fringe-stopping interferometer ($\tau_0=w_p/c$), namely, from phase tracking in the pointing direction. This is a near-field version of the traditional phase delay term for astronomical sources that is present in Equation \eqref{eq:ast_vis}. It assumes spherical wavefronts with a direct path of propagation between the source and receiver.

\subsubsection{RFI signal model}\label{sec:rfi_signal}

The complex-valued RFI signal at each antenna is modelled using a GP over time using inducing points $A_p(\bm{t}')$. The modelled RFI signal represents the intrinsic signal emitted by the RFI source in the direction of the antenna, including any signal corruptions along the way, until it is measured in the antenna by the receiver, as well as any position errors of the source or receiver. Therefore, signal corruptions such as Faraday rotation, ionospheric effects, the antenna voltage pattern, and the antenna gains are all included in the RFI signal $A_p(t)$. For a comprehensive overview of possible signal corruptions see \citet{Smirnov2011b}.

We use inducing points as described in Section \ref{sec:interpolation} to reduce the number of parameters in our model based on the expected variability of the RFI signal. Typically, the deciding factor in the RFI signal variability is the movement of the RFI source through the antenna primary beam sidelobes. Analogously to the gains, the RFI signal inducing points are interpolated to the desired sampling points with
\begin{equation}
    A_p \left( \bm{t} \right) = \bm{I}^A A_p \left( \bm{t}' \right),
\end{equation}
where the locations $\bm{t}$ are determined by satisfying Equation \eqref{eq:rfi_bounds}. The prior distribution of the RFI signal at each antenna is
\begin{equation}
    p \left( A_p \left( \bm{t}' \right) \right) = \mathcal{N} \left( \bm{0}, \bm{C}_A' \right),
\end{equation}
where $\bm{C}_A'$ is the covariance calculated from the SE covariance function evaluated at the inducing points $\bm{t}'$. Therefore $\bm{C}_A'(\sigma^2_\text{RFI}, \tau_\text{RFI})$ is an $M\times M$ real-valued matrix where $M$ is the number of inducing points. $\sigma^2_\text{RFI}$ can be set to the level at which the telescope response becomes non-linear. If the RFI signal is strong enough to push into the non-linear regime of the telescope, this method will not work in its current formulation. Currently, TABASCAL assumes linearity, however, the non-linearity could in future be modelled. 

Assuming the fastest variation in the RFI signal is due to the movement through the primary beam sidelobes, then $\tau$ value can be chosen based on the apparent sidelobe traversal rate. To approximate this, we divide the approximate angular sidelobe spacing, $\frac{\lambda}{D}$, by the apparent angular velocity, $\frac{|\vec{v}_\text{RFI}|}{R_\text{RFI}}$, of the RFI source, and further divided by four, to give four samples per cycle. Therefore,
we have\begin{equation}\label{eq:rfi_l}
    \tau_\text{RFI} \approx \frac{\lambda R_\text{RFI}}{ 4 D |\vec{v}_\text{RFI}|},
\end{equation}
where $D$ is the dish diameter. 

Given the combination of a geometric delay term and use of a GP to model the RFI signal at an antenna, this can be thought of as a fringe rate filter, relative to the array centre, in the direction determined by the geometric delay term. Since we use the SE covariance function, this corresponds to an SE power spectrum with $\eta_0 = 1/2\pi \tau$. Effectively, this fringe stops in the expected direction of the RFI source and applies a fringe filter determined by the GP. Following this logic we can see that anything that causes a differential fringe rate, from this direction, can be effectively modelled by the GP, for that particular source. A differential fringe rate can be caused by many things in the signal chain including the variability of the RFI signal at emission (duty cycle), estimated position errors, ionospheric fluctuations in the direction of the source, beam attenuation from the source's antenna, as well as, the telescope primary beam as is accounted for in Equation \eqref{eq:rfi_l}.

\subsubsection{Computational considerations}

In the previous sections, we  describe the GP models used for the astronomical visibilities, the antenna gains, and the RFI signals. Two particular types of GP models were used, namely the Fourier-based method and the inducing points method. These particular GP formulations are not requirements of our method, however, they were chosen with computational considerations in mind. 

Since the astronomical visibilities vary on different time scales, ideally, a separate prior covariance is used for each baseline. Unfortunately, this would lead to unfavourable scaling when storing and/or calculating the astronomical visibility covariances. The benefit of restricting ourselves to GPs with covariance functions that are diagonal in Fourier space, is specifically in terms of computational and memory requirements. Given $N_T$ time steps at which we want to infer the signal, a general GP requires us to calculate $\mathcal{O}(N_T^2)$ terms forming the covariance matrix. Inverting this matrix would have $\mathcal{O}(N_T^3)$ computational cost. Since $\bm{F}$ can be applied via the Fast Fourier Transform (FFT) and $\bm{P}_{\eta\eta}$ is diagonal, the inverse of $\bm{C}_{tt}$ in this case can be computed in $\mathcal{O}(N_T\log N_T)$ with only $\mathcal{O}(N_T)$ memory requirements.

For the RFI signal and antenna gains, the inducing points GP method was used. Since the expected behaviour of these components is roughly the same on each antenna, we are able to use the same prior covariance across the antennas and subsequently the same interpolator as defined in Equation \eqref{eq:interpolator}. There is little additional value in changing to the Fourier-based GP method as the number of covariance matrices to compute (and potentially store) is independent of the number of antennas/baselines. This could of course change if we are dealing with, for example, a particularly unstable antenna, or a more turbulent ionosphere above a certain set of antennas.

As can be seen from Equation \eqref{eq:interpolator}, the calculation of an interpolation matrix involves the inversion of a real-valued matrix of size $N \times N$, where $N$ is the number of known locations, $\bm{t}'$, one wants to interpolate from. The real-valued interpolator matrix $\bm{I}_{tt'}$ has a size of $M \times N$, where $M$ is the number of points to interpolate to, $\bm{t}$. It is therefore computationally favourable to reduce $N$ as far as possible without negatively impacting the final solution. A reasonable choice is $N = \Delta T / \tau$ where $\Delta T$ is the interval over which to interpolate and $\tau$ is the correlation length used in the SE covariance function. 

For the antenna gains, the number of interpolation locations (inducing points) can be kept very low and is typically kept at $N=3$ for a given target observation block of around 15 minutes. However, for the RFI signal, the number of inducing points $N$ and interpolated values $M$ could become very large. If this is the case, it becomes favourable to use a Fourier-based GP model with some adaptations to reduce edge effects and parameter count. We do not investigate this alternative here, however, we do note its potential to improve the scaling of this method in unfavourable scenarios.  

\subsection{Posterior approximation}\label{sec:post_approx}

In Section \ref{sec:forward_model}, the different components that make up our Bayesian model are defined. This includes the likelihood and all of the prior terms which are assumed to be independent of one another. The full prior distribution is therefore simply the product of the individual prior distributions. The posterior distribution, calculated with Equation \eqref{eq:bayes}, gives us the updated distribution over our model parameters, after the inclusion of information supplied by our data. The central challenge after defining our probabilistic model becomes estimating the posterior distribution.  

There are a number of ways to estimate the posterior distribution. The most rigorous is to use a Markov chain Monte Carlo (MCMC) scheme such as Hamiltonian/hybrid Monte Carlo \citep{Brooks2011}, where samples can be drawn directly from the posterior. However, MCMC techniques can be slow to converge and often infeasible in high-dimensional settings such as for our problem. Alternatives include: variational inference \citep{Blei2017}, the Laplace approximation \citep{Tierney1986}, and the simplest, approximating with a delta distribution, namely, a maximum a posteriori (MAP) estimate only. In this work, we stick to MAP estimation for computational reasons. The Laplace approximation approximates our posterior with a Gaussian centred on the MAP point. We have implemented a scalable method to estimate the posterior using the Laplace approximation, which can be optionally run after the optimisation step.

Our posterior approximation method is therefore summarised as using a non-linear optimisation routine to estimate the MAP point and thereafter, optionally, estimating the posterior covariance from some notion of the posterior information. In this section, we describe standardised coordinates which help to decorrelate our parameter space and serve to precondition our optimisation routine, and then we describe the method used to estimate the posterior covariance in a scalable way.

\subsubsection{Standardised coordinates and optimisation}\label{sec:std_coords}

Standardised coordinates \citep{knollmuller2019metric} are a coordinate system in which the prior distribution follows a simple and uncorrelated model. Most commonly, this is a standard normal distribution. This is also referred to as a non-centred parametrisation \citep{Betancourt2015}. In a hierarchical model, the prior distribution itself is also parametrised, in this case, it is referred to as the reparametrisation trick \citep{Kingma2013AutoEncodingVB}. In this work, we do not employ a hierarchical model, however, our method still benefits from a standardised parametrisation given the correlations in our prior. Given a prior distribution, $\mathcal{N}(\bm{\mu}, \bm{\Sigma})$, over a set of parameters $\bm{\Psi}$ we can define new standardised parameters $\tilde{\bm{\Psi}}$ as 
\begin{equation}\label{eq:std_coords}
    \tilde{\bm{\Psi}} = \bm{L}^{-1} \left( \bm{\Psi} - \bm{\mu} \right),
\end{equation}
where $\bm{L}$ is a matrix square root of $\bm{\Sigma} = \bm{L}\bm{L}^T$. Therefore, the prior distribution over the standardised parameters is 
\begin{equation}
    p \left( \tilde{\bm{\Psi}} \right) = \mathcal{N} \left( \bm{0}, \bm{1} \right),
\end{equation}
where $\bm{0}$ is the vector with all zeros and $\bm{1}$ is the identity matrix, i.e. the prior on $\tilde{\bm{\Psi}}$ is the standard normal distribution. $\tilde{\bm{\Psi}}$ can now be used as the base parameters of our model and all parameters described in Section \ref{sec:forward_model} can be calculated using
\begin{equation}
    \bm{\Psi} = \bm{\mu} + \bm{L} \tilde{\bm{\Psi}}.
\end{equation}

Since our model makes heavy use of GP priors, our parameter space is strongly correlated in the space where the prior dominates. This makes both sampling and optimisation in such a space very inefficient due to slow convergence. This is often aided by preconditioning the optimisation routine. By using standardised coordinates, we achieve the same objective. Both methods are equivalent when the pre-conditioner is $\bm{L}^{-1}$, however, by explicitly doing this ourselves we have the option to use many out-of-the-box optimisation routines that are available. In our work, we have used the AdaBelief optimiser \citep{zhuang2020adabelief}.


\subsubsection{Covariance estimation}\label{sec:cov_est}

The standard way of performing a Laplace approximation is to calculate the posterior Fisher information, which is defined as
\begin{equation}\label{eq:fisher}
    \bm{\mathcal{I}}(\bm{\theta}) = \mathbb{E} \left[ \left( \frac{\partial}{\partial \bm{\theta}} \log p\left( \bm{\theta} | \mathcal{D} \right) \right)^2 \right],
\end{equation}
and then inverting the resulting matrix to get the covariance matrix. In Equation \eqref{eq:fisher} above, $p\left( \bm{\theta} | \mathcal{D} \right)$ is the posterior distribution over model parameters $\bm{\theta}$. In a model like ours, where the number of parameters $N_P$ is on the order of $10^5$ - $10^6$, this is infeasible due to the computational cost of inversion being $\mathcal{O}(N_P^3)$ and memory requirements of $\mathcal{O}( N_P^2) \approx 1$ TB. An alternative method is to draw samples from the Gaussian distribution with the desired covariance. To do this we employ the method of Gaussian constrained realisations \citep{Hoffman1991}. We derive this method for the non-linear case in Appendix \ref{app:cov_est}. The main result is
\begin{align}\label{eq:cov_samp}
    \Delta\bm{\theta} &= \hat{\bm{\Sigma}} \left( \bm{J}^T \bm{\Sigma}_N^{-1} \Delta\bm{\psi} + \bm{\Sigma}_\Pi^{-1} \Delta\bm{\phi} \right),
\end{align}
where $\Delta\bm{\theta}$ are the perturbations about the MAP point obtained from optimisation. $\hat{\bm{\Sigma}}$ in Equation \eqref{eq:cov_samp} is defined as $\bm{\mathcal{I}}^{-1}(\bm{\theta})$ and applied in the equation in an implicit manner so that is is never directly evaluated in its entirety for the reasons given above. We also have  
\begin{align}
    \Delta \bm{\psi} \sim \mathcal{N}(\bm{0}, \bm{\Sigma}_N), \quad \text{and} \quad
    \Delta \bm{\phi} \sim \mathcal{N}(\bm{0}, \bm{\Sigma}_\Pi),
\end{align}
where $\bm{\Sigma}_N$ is the visibility noise covariance and $\bm{\Sigma}_\Pi$ is the prior covariance. Fortunately, both of these distributions can be trivially and scalably sampled due to their diagonal covariances in standardised coordinates. Obtaining the posterior covariance estimate is then done as,
\begin{equation}
    \hat{\bm{\Sigma}} = \mathbb{E} \left[ \Delta\bm{\theta} \Delta\bm{\theta}^T \right].
\end{equation}
Since $\hat{\bm{\Sigma}}$ has such a high dimensionality, this method allows us to efficiently evaluate marginalised covariances without evaluating the entire posterior covariance matrix.

\section{Data simulations and evaluation set}\label{sec:sims}

In this section we briefly describe the simulation set used to analyse the performance of TABASCAL across a wide range of RFI strengths. We also describe our comparison methods used to compare against TABASCAL's performance. Most notably, the uncontaminated data which serves as a benchmark for statistically perfect signal separation.   

\subsection{Simulation set description}\label{sec:sim_desc}

\begin{figure}
  \centering
  \includegraphics[width=0.5\textwidth]{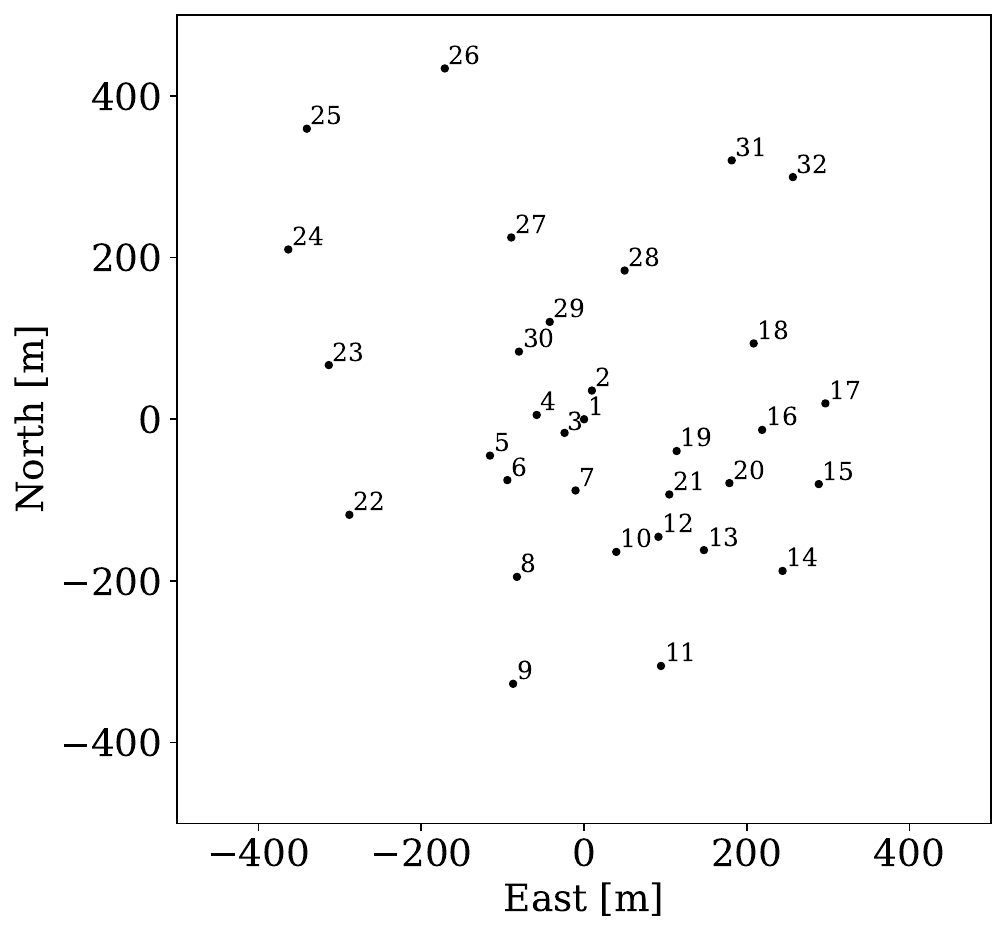}
  \caption{Antenna layout of the 32 MeerKAT antennas used in our simulations.}
  \label{fig:ants}
\end{figure}

To evaluate TABASCAL rigorously, we tested on 92 simulated observations with widely varying RFI signal amplitudes. We have simulated observations with a 32 antenna (from the core) MeerKAT \citep{Jonas2018} array and a single frequency channel. Figure \ref{fig:ants} shows the antenna layout used in our simulations. Each observation is 15 minutes long with two second integration times. A summary of the telescope parameters used is shown in Table \ref{tab:tel_sim} in Appendix \ref{sec:parameters}. The gains were set to vary linearly in time with starting values for each antenna drawn from $\mathcal{N}(1, 0.05^2)$ and $\mathcal{U}(-90^\circ, 90^\circ)$, for the amplitudes and phases respectively. $\mathcal{U}(-90^\circ, 90^\circ)$ represents the uniform distribution between the bounds $-90^\circ$ and $90^\circ$.

A detailed description of the simulations used in this work is provided in \citetalias{Finlay2023}. The only differences in this work are:
\begin{enumerate}
    \item the use of 32 antennas instead of 64 for computational and storage reasons;    \item the change of astronomical flux distribution from an exponential to a power law;
    \item the use of TLE-based satellite trajectories as opposed to a simple circular orbit.
\end{enumerate}
To accurately simulate the contribution from RFI sources we use a geometric delay term, as given in Equation \eqref{eq:rfi_phase}, which considers RFI sources to be in the near-field for the interferometric array. Additionally, we include the effect of fringe winding due to the non-sidereal movement of the RFI source through the integration used in Equation \eqref{eq:rfi_vis} where the required sampling is determined by Equation \eqref{eq:rfi_sampling}. Finally, in \citetalias{Finlay2023}, it was determined all RFI sources outside of the telescope site itself would be in the far-field of a single antenna where the primary beam term for astronomical sources is valid. However, due to RFI sources being in the near-field on the array level, the angle of signal arrival is different for each antenna and, as such, included in the simulations. The intrinsic RFI signal is assumed to be constant in time in the simulations. This will not be true in real data as the RFI source has its own emission pattern and duty cycle. However, in our Bayesian forward model, we allow for this effect by modelling the RFI signal ($A_{ps}(\bm{t})$), per source ($s$), and per antenna ($p$) to be time varying with a user tunable parameter for its variation.

The included satellite-based RFI sources are selected from the GPS satellites that passed the target direction within 45 degrees. This corresponded to between two and nine satellites for each 15 minute simulation. Each observation has 100 point sources uniformly distributed within the field of view (1.42 degrees) with minimum angular separations of 180". The theoretical synthesised beam size is around 70" for an uncontaminated dataset. The source fluxes are drawn from a power-law distribution given by 
\begin{equation}
    p(S) \propto S^{-\beta},
\end{equation}
where $\beta=1.6$ \citep{Intema2011} and, minimum and maximum fluxes set to $10\sigma_I = 14$ mJy and 1 Jy respectively. $\sigma_I$ is the theoretical image noise for an RFI-free dataset with visibility noise of $\sigma_n = 0.65$ Jy per 2 second integration time, as is used in our simulations. The mean astronomical visibility amplitude is 1.1 Jy and the mean RFI visibility amplitude varies from $10^{-3}$ Jy to $10^3$ Jy.

Each observation uses an independent sky, gain, and noise realisation. The RFI power was artificially varied to give a large range in RFI visibility amplitudes for testing purposes. For the simulations used in this paper, the maximum fringe frequency, $\nu_f$, of all satellites was approximately 0.6 Hz. From Equation \eqref{eq:rfi_bounds}, we calculate the maximum S/N for our sampling rate to be $4.4 \times 10^5$ and a minimum S/N of 260. 

We found TABASCAL is capable of successfully removing RFI up to an S/N of $9.4 \times 10^3$ equating to a mean RFI S/N of $2.2 \times 10^3$. However, this is on simulations using a limited sampling rate due to compute constraints. 

\subsection{Evaluation set description}\label{sec:eval_desc}

To analyse the performance of TABASCAL over a range of RFI strengths, we have binned the simulations over RFI S/N into five bins. The RFI S/N is calculated as the average RFI visibility amplitude across all data points divided by the standard deviation of the visibility noise. The average RFI visibility amplitude is defined as
\begin{equation}
    |V^\text{RFI}| = \frac{1}{N_D} \sum_{i=1}^{N_D} |V_i^\text{RFI}|,
\end{equation}
and $|V^\text{AST}|$, the average astronomical visibility amplitude, is similarly defined. Therefore the RFI visibility S/N is defined as 
\begin{equation}
    \text{S/N}(|V^\text{RFI}|) = \frac{|V^\text{RFI}|}{\sigma_n}.
\end{equation}

We compare TABASCAL to three other cases: (i) `uncontaminated', (ii) `perfect $3\sigma$ flagging', and (iii) AOFLAGGER. The perfect $3\sigma$ flagging and AOFLAGGER cases use visibilities that have been correctly calibrated; namely, the true calibration solutions have been applied to the uncalibrated, simulated visibilities. The resulting data is then flagged for RFI using either a perfect $3\sigma$ flagging or AOFLAGGER. Flagged data is not used in any downstream processing. This is where TABASCAL is able to gain an advantage as it does not flag the data but rather corrects it so that more data is available for downstream processing. Perfect $3\sigma$ flagging, is referred to as such because the true calibration solutions have been applied and the flagging has had access to the true (noise-free) astronomical visibilities. This is obviously not a realistic scenario but is included to show the limits of flagging as a method to address the removal of RFI as compared to using TABASCAL, where the visibilities are corrected and therefore no flagging is applied. 

Perfect $3\sigma$ flagging is where we have flagged based on the difference between the true astronomical visibilities, $V^\text{AST}$, and the correctly calibrated visibilities, $V^\text{CAL}$. We have flagged where the amplitude of this is greater than three times the true noise amplitude, $\sigma_n$. Mathematically, this is represented as 
\begin{equation}
    \text{Flag}_i = |V_i^\text{CAL} - V_i^\text{AST}| > 3 \sigma_n.
\end{equation}

The AOFLAGGER runs show a slightly more realistic situation. As with the perfect $3\sigma$ flagging, we use the correctly calibrated visibilities but then the AOFLAGGER algorithm has been applied. We make use of three passes on the data, each with a dedicated strategy taken directly from CARACAL  \citep{Jozsa2020}, a radio interferometry data-reduction pipeline software \footnote{\href{https://github.com/caracal-pipeline/caracal/tree/master/caracal/data/meerkat_files}{CARACAL }'s AOFLAGGER strategies.} . This shows a more realistic scenario, however, the strategies used have not been optimised for our dataset. Although they are for generic MeerKAT data, which is what we have simulated, except for one channel only.

The uncontaminated case refers to the sum of the true astronomical visibilities and the same noise realisation used in the other RFI contaminated cases. This stands as our reference point and should be the limit that can be reached by TABASCAL when it is working perfectly and no significant amount of information about the astronomical visibilities has been given. 

From the 92 simulations, 86\% (79) resulted in a $\bar{\chi}^2$ < 1.1. $\bar{\chi}^2$ is the $\chi^2$ per data point and is defined as
\begin{equation}
    \bar{\chi}^2 = \frac{1}{N_D} \sum_i^{N_D} \left(\frac{V_i^\text{OBS} - \hat{V}_i}{\sigma_n} \right)^2,    
\end{equation}
where $N_D$ is the number of real-valued data points, $V_i^\text{OBS}$ are the data, $\hat{V}_i$ are the estimated visibilities, and $\sigma_n$ is the noise standard deviation. The visibilities are split up into the real and imaginary components here. 

The following results only consider these 79 simulations where TABASCAL successfully converged according to this criteria. In the next section, we show that TABASCAL recovers astronomical visibilities comparable in accuracy to the situation where an idealised telescope has observed the same sky with no RFI contamination, namely, the uncontaminated case, effectively allowing us to `see through' the satellites in the contaminated observation.

Table \ref{tab:prior_params} shows the prior parameters that have been used throughout the results presented in Section \ref{sec:results}. These priors are chosen either to be non-informative, based on expectations of the signal due to telescope and observation considerations in the case of the RFI and astronomical signals, or based on calibration observation information for the gains. The astronomical signal prior is described in Section \ref{sec:ast_vis_model}, the RFI signal prior is described in Section \ref{sec:rfi_signal}, and the prior on the gains is described in Section \ref{sec:gain_model}.

\section{Results}\label{sec:results}

In this section, we discuss the performance of TABASCAL in terms of the astronomical visibility recovery (Section \ref{sec:vis}), the resulting image quality (Section \ref{sec:imaging}) and subsequent point source recovery statistics (Section \ref{sec:pnt_src}). We compare the performance of TABASCAL to the uncontaminated ideal case, a perfect $3\sigma$ flagging and calibration situation, and finally to correctly calibrated data that was then flagged using AOFLAGGER \citep{offringa2010aoflagger}. We finish this section with an analysis of the gain phase calibration capabilities of TABASCAL by leveraging the RFI S/N in Section \ref{sec:phase_cal}.

\subsection{Visibilities}\label{sec:vis}

\begin{figure}
  \centering
  \includegraphics[width=0.5\textwidth]{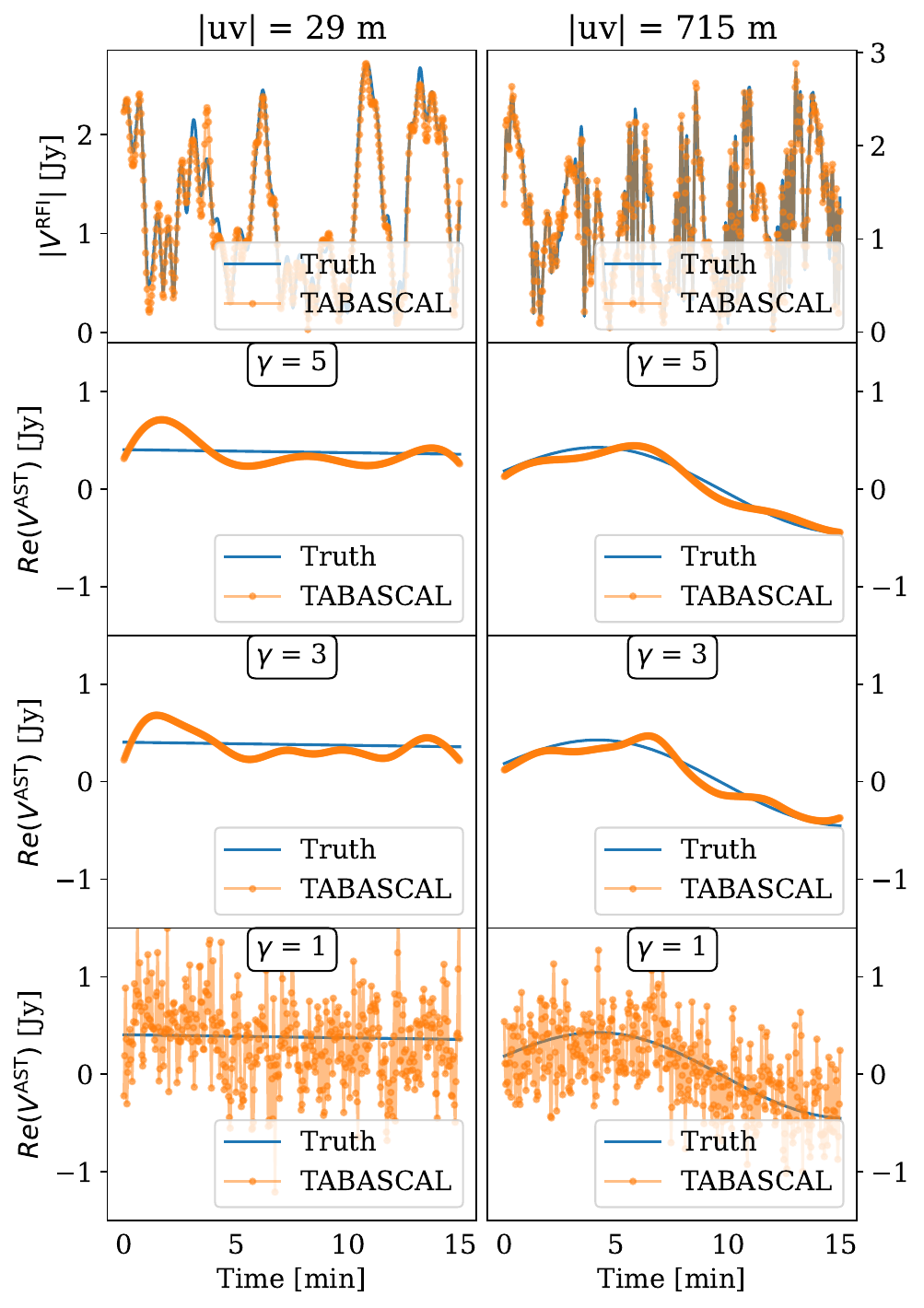}
  \caption{Example predictions from TABASCAL for two different baselines (left and right) on an observation containing six satellites. The orange dotted curve shows the TABASCAL prediction and the blue curve shows the true value. The top panel shows this for the RFI visibility magnitude and the lower panels show the real part of the astronomical visibility. The top panel corresponds to a TABASCAL run where $\gamma=5$, as defined in Equation \eqref{eq:power_spec}. The three lower panels show the results when varying the prior parameter $\gamma$ for the astronomical visibilities. Finally, $\gamma$ controls the smoothness of the solutions.}
  \label{fig:vis_ex}
\end{figure}

Due to the presence of multiple RFI sources with differing fringe rates, the resulting  RFI visibility signal is non-trivial. In Fig. \ref{fig:vis_ex}, we show an example of the TABASCAL prediction for the RFI visibility magnitude and real part of the astronomical visibility for two distinct ($uv$) baselines, a short (29 m) and a long baseline (715 m). For the astronomical visibility prediction, in the lower three panels, we have changed the prior hyperparameter $\gamma$ which controls the smoothness of the estimated astronomical visibilities. Equation \eqref{eq:power_spec} gives the functional form of the prior covariance in fringe rate space, namely, the power spectrum. Here, $P_0$ has been fixed to $10^7$ Jy$^2$, corresponding to a prior standard deviation of $\sigma^\text{AST}\approx 7$ Jy in the time domain (real space). The mean astronomical visibility magnitude is approximately 1 Jy across all observations and baselines. We found no significant effect on the solutions when varying this hyperparameter unless it was chosen to be too small. The main consideration when choosing this hyperparameter should be to encompass the expected signal, namely, $\sigma^\text{AST} > 1$ Jy for our simulations. We expect a reasonable estimate can be found from a neighbouring uncontaminated channel. 

To be noted is the difference in time variations of the true signals between the left and right hand side panels (short and long baselines respectively). From the lower three/six panels, we observe the effect of varying $\gamma$ on the TABASCAL predicted astronomical visibilities. As $\gamma$ is increased, the predicted visibilities become smoother and have lower variance. The trade-off here becomes increased correlation between samples. We found that the subsequent imaging results, in Section \ref{sec:imaging}, showed equal image noise for all values of $\gamma$ tested which varied between 1 and 5. As can be seen from the lowest panel in Fig. \ref{fig:vis_ex}, $\gamma=1$ leads to the astronomical visibilities starting to fit the noise, this can also lead to residual RFI signal leaking into the astronomical visibilities. As $\gamma \rightarrow \infty$, the power spectrum (fringe filter), as defined in Equation \eqref{eq:power_spec}, becomes Gaussian. This is the desired result, however, this leads to numerical instabilities due to the variance becoming 0 for large $\eta$ values. As such, the recommendation is to make $\gamma$ as large while avoiding these numerical instabilities. Throughout the rest of these results we use $\gamma=5$.

For all of these solutions $\eta_0$ is kept fixed at 1 mHz, and is chosen according to Equation \eqref{eq:ast_fringe} which is dependent on the telescope configuration and the expectation of no strong off-axis sources. $\eta_0$ can have a unique value for each baseline, as calculated using Equation \eqref{eq:ast_fringe}. However, due to the fringe period being much longer than our 15 minute observations for many baselines, this would lead to mostly constant solutions on the shorter baselines. Therefore, we have chosen $\eta_0$ to correspond to one of the longer baselines. We note that using a unique $\eta_0$ as calculated using Equation \eqref{eq:ast_fringe} leads to comparable results and is expected to be the better choice when using TABASCAL on longer observations.  

In the top panel of Fig. \ref{fig:vis_ex}, we can clearly see the accuracy of the RFI prediction on a rather complicated signal from the combination of six satellites. We have chosen a simulation where the RFI S/N is around 2 to show its performance in the lower-mid S/N range. When looking at solutions for an S/N of >10, the errors are no longer visible. This is because the prediction error is constant and does not scale with the RFI strength. 

\begin{figure}
  \centering
  \includegraphics[width=0.5\textwidth]{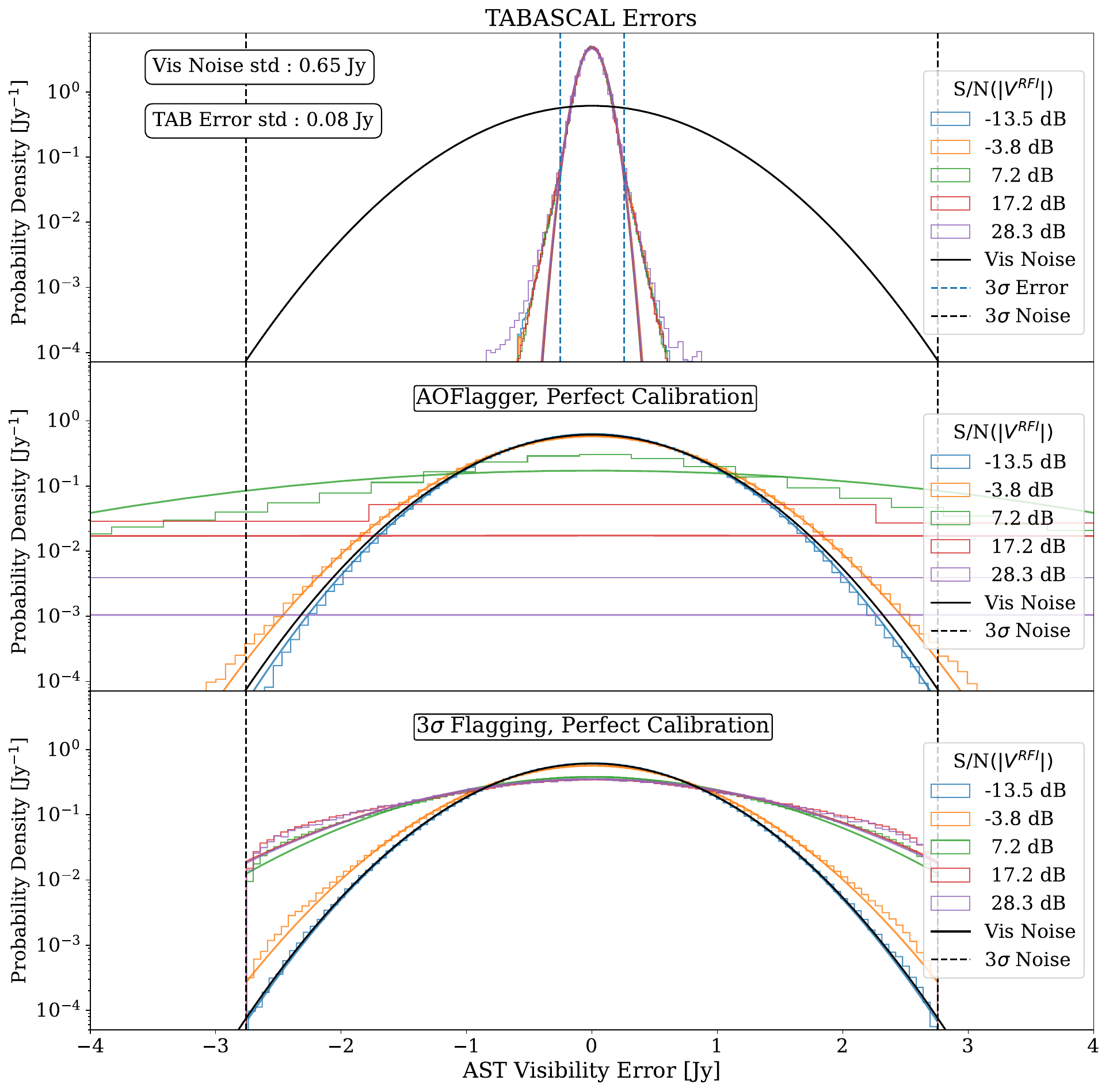}
  \caption{Errors in the astronomical visibility predictions from TABASCAL and the other cases for comparison as the S/N of the RFI is varied. The black curve shows the distribution of the visibility noise used in the observations and therefore corresponds to the errors in the uncontaminated case. In each panel five bins of RFI strength are used where multiple observations are bundled together. The coloured curves show the Gaussian fit to the error distributions. (i) The top panel shows the errors in the TABASCAL predicted visibilities. (ii) The middle and bottom panels show the errors from the AOFLAGGER and perfect $3\sigma$ flagging cases where flagged data is not included in the histograms.}
  \label{fig:tab_vis_error}
\end{figure}
\FloatBarrier

In the top panel of Fig. \ref{fig:tab_vis_error}, we show the error distribution in the astronomical visibility prediction from TABASCAL. The individual errors, $\epsilon_i$, are defined as 
\begin{equation}\label{eq:ast_error}
    \epsilon_i = \hat{V}^\text{AST}_i - V^\text{AST}_i,
\end{equation}
where $\hat{V}^\text{AST}$ is the TABASCAL prediction. The real and imaginary parts are concatenated. The black curve represents the distribution for the uncontaminated case where there is only thermal Gaussian noise added. We see that the error distribution from TABASCAL predicted astronomical visibilities has Gaussian errors down to $3\sigma$ relative to their fitted distributions. The $3\sigma$ error limit is shown as dashed vertical blue bars. We also see that the error distribution is almost identical at all RFI scales except for the largest bin. This is likely because of the limited sampling rate used in our RFI simulations. The accuracy of our RFI simulations diminishes at higher S/N due to this limited sampling rate in our simulations when averaging the sub-integration samples. Section \ref{sec:rfi_vis} describes these bounds in detail. We only have theoretical guarantees of simulation accuracy up to RFI S/N of 260. Unfortunately, data volumes above the sampling frequency used would have led to a reduced number of simulations being possible. These deviations from Gaussianity appear to have little effect on the subsequent imaging and point source recovery results.

In the middle panel of Fig. \ref{fig:tab_vis_error}, we see that AOFLAGGER does not manage to flag much of the low level RFI contaminated visibilities. This is likely due to the lack of an uncontaminated reference for the algorithm as we only include a single contaminated frequency channel in our simulations. For the perfect $3\sigma$ flagging case in the bottom panel, we see the hard $3\sigma$ flagging threshold (with respect to the noise distribution) used and note the increased number of very low level visibility errors below the threshold. For both the flagging cases, we see that the error distribution follows very closely to that of the uncontaminated case for RFI strengths below an S/N of 1 as would be expected. 

Of note is the reduced error variance for the TABASCAL predicted visibilities vs the uncontaminated case. As can be seen already from Fig. \ref{fig:vis_ex}, we are able to trade off error variance with increased correlation over time in our solutions through the variation in the $\gamma$ parameter in the prior. Due to this trade-off, we find that subsequent imaging and point source recovery results remain stable. This is due to the gridding step in imaging which uses a convolution kernel to resample the visibilities onto a grid. We are effectively applying a convolution in the time axis through our astronomical visibility prior.

\subsection{Imaging}\label{sec:imaging}

\begin{figure}
  \centering
  \includegraphics[width=0.45\textwidth]{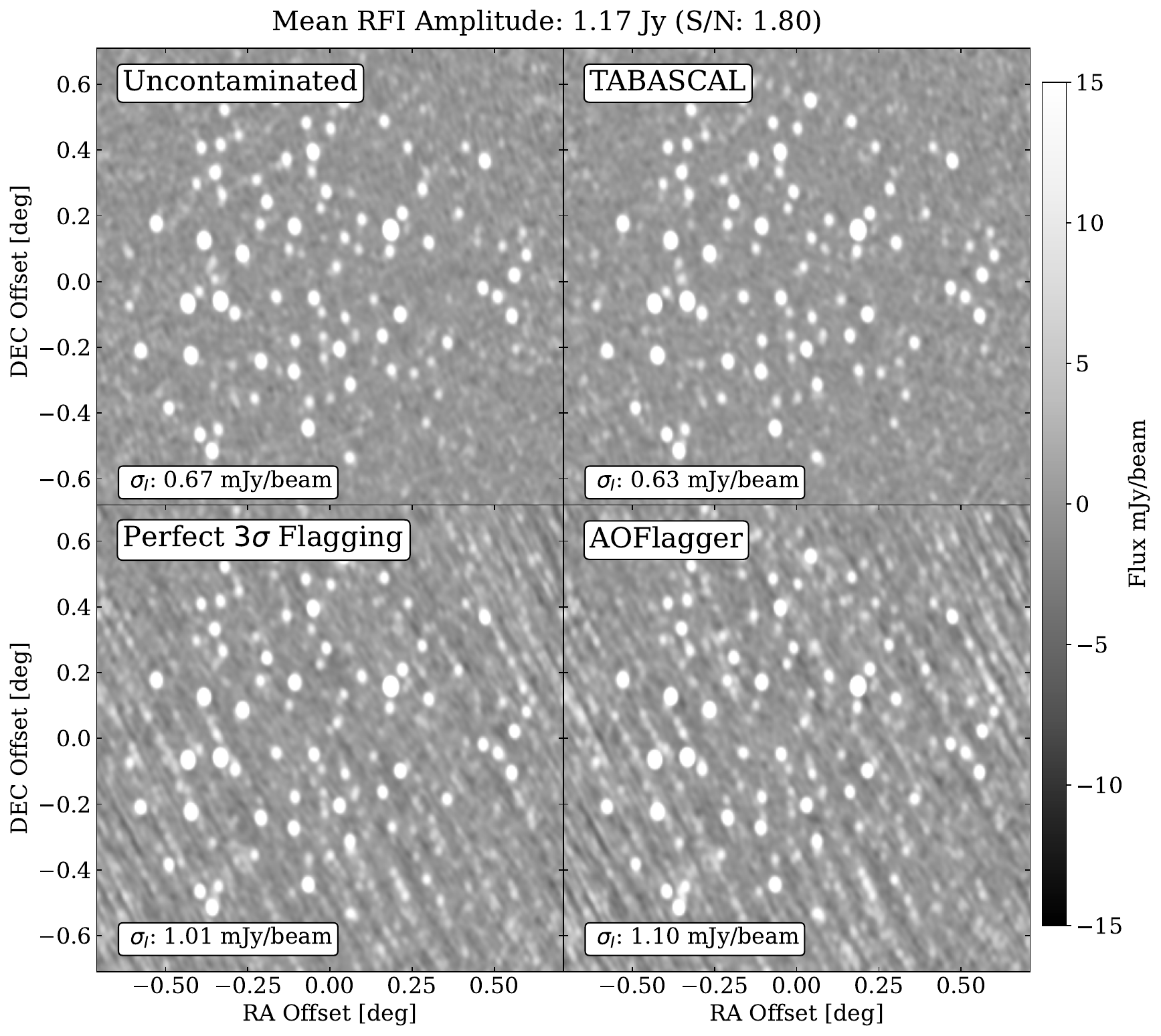}
  \caption{Images constructed from the same observation with our four different cases. Top Left: No RFI contamination. Top Right: TABASCAL fully removes the RFI and recovers the astronomical signal with comparable image noise to the uncontaminated data. Bottom Left and Right: The images from perfect $3\sigma$ flagging and AOFLAGGER respectively, showing significant striping due to residual RFI contamination. The mean RFI S/N in this data was 1.8, showing that significant issues occur for traditional flagging methods even at weak RFI. Here, a perfect $3\sigma$ flagging means that any RFI with true amplitude greater than $3\times$ the noise is perfectly removed.}
  \label{fig:img_comp1}
\end{figure}

\begin{figure}
  \centering
  \includegraphics[width=0.45\textwidth]{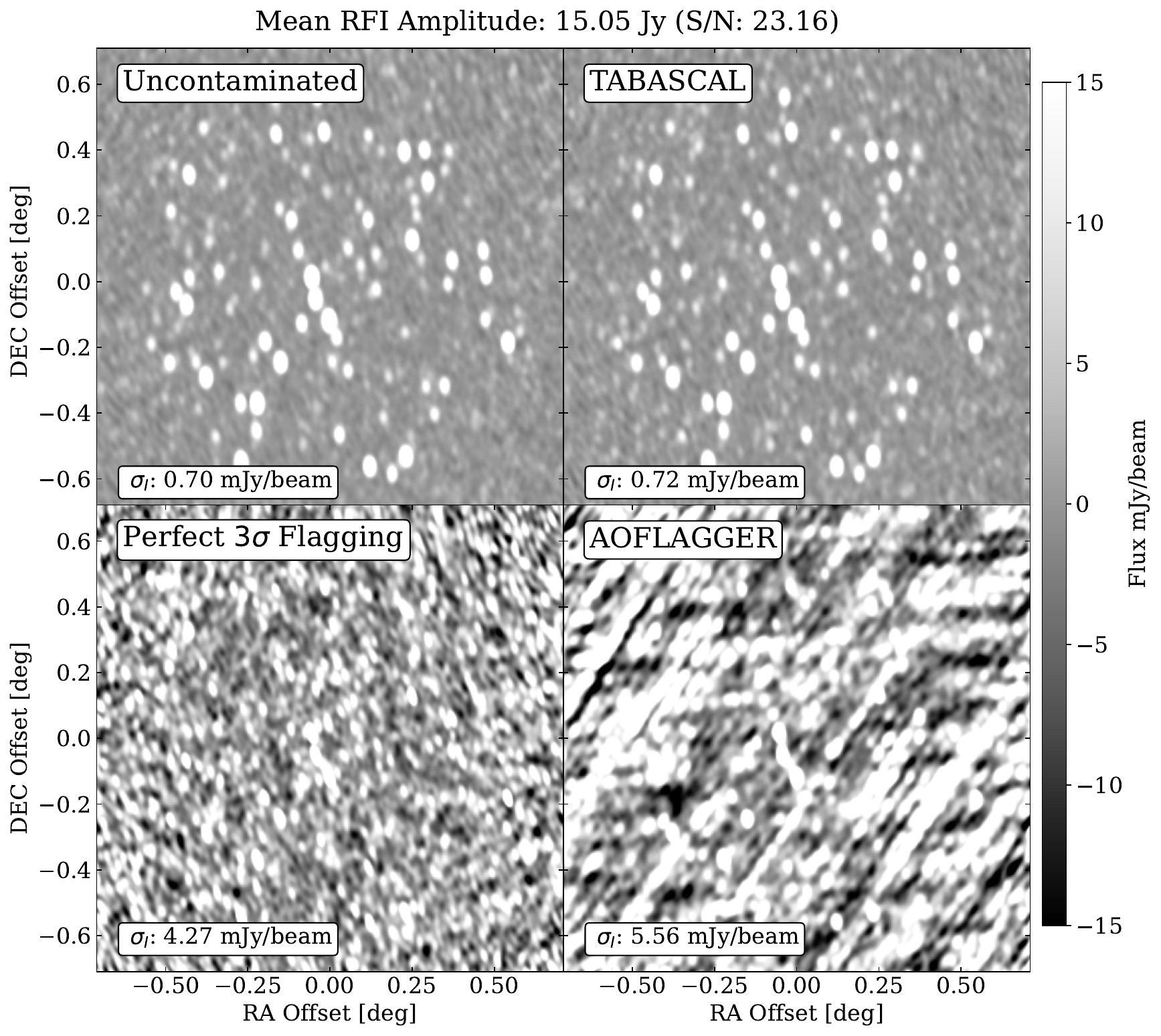}
  \caption{Images constructed from the same observation with our four different cases. Top Left: No RFI contamination. Top right: TABASCAL fully removes the RFI and recovers the astronomical signal with comparable image noise to the uncontaminated data. Bottom-left and right: The images from perfect $3\sigma$ flagging and AOFLAGGER  respectively, showing significantly higher image noise of 6x and 8x respectively. The AOFLAGGER image shows significant striped image artefacts which largely invalidates its use for science, with purity and completeness of around $20\%$ and $50\%$ respectively (see Figures \ref{fig:purity} and \ref{fig:completeness}). The mean RFI S/N in this data was 23. For large RFI amplitudes the quality differences between TABASCAL and the flagging methods increases further.}
  \label{fig:img_comp10}
\end{figure}

TABASCAL is used to estimate the uncontaminated astronomical visibilities, however, one of the main data products from radio interferometry are images. To evaluate the performance of TABASCAL in this regard, the visibilities from all simulated observations for all four cases have been imaged using WSCLEAN\citep{2014MNRAS.444..606O}. The same imaging parameters \footnote{\texttt{wsclean -size 426 426 -scale 12asec -pol xx -weight natural -niter 1000000 -magin 0.3 -auto-mask 1.0 -auto-threshold 0.3 -no-negative observation.ms}} were used for all cases and observations. In Figures \ref{fig:img_comp1} and \ref{fig:img_comp10}, we show an example set of images for all four cases being compared. We have used an observation where the mean RFI visibility amplitude is 1.2 \& 15 Jy corresponding to S/N s of 1.8 and 23. These images were chosen based purely on S/N and therefore should be a representative sample of other results at similar S/N values. 

TABASCAL shows very comparable image quality to the uncontaminated case at both S/N levels. In contrast, the flagging cases are showing higher image noise and image artefacts. At RFI S/N of 1.8, visible image artefacts of diagonal banding are present for both cases, as well as, more than 50\% higher image noise. These become especially apparent at a higher S/N of 23 where the AOFLAGGER image becomes unusable with strong diagonal striping and an image noise eight times higher than TABASCAL. At this S/N level, even the perfect $3\sigma$ flagging case has six times higher image noise. This really shows the limits that flagging as a mitigation strategy has, and in all likelihood, this data would not even be included in any downstream data reduction. A multi-frequency flagging strategy would most likely flag all of this data resulting in no possible image in this particular frequency channel. In contrast, TABASCAL shows effectively equivalent image quality to the uncontaminated case and even outperforms it in the specific example shown for the low S/N case.  

\begin{figure}
  \centering
  \includegraphics[width=0.45\textwidth]{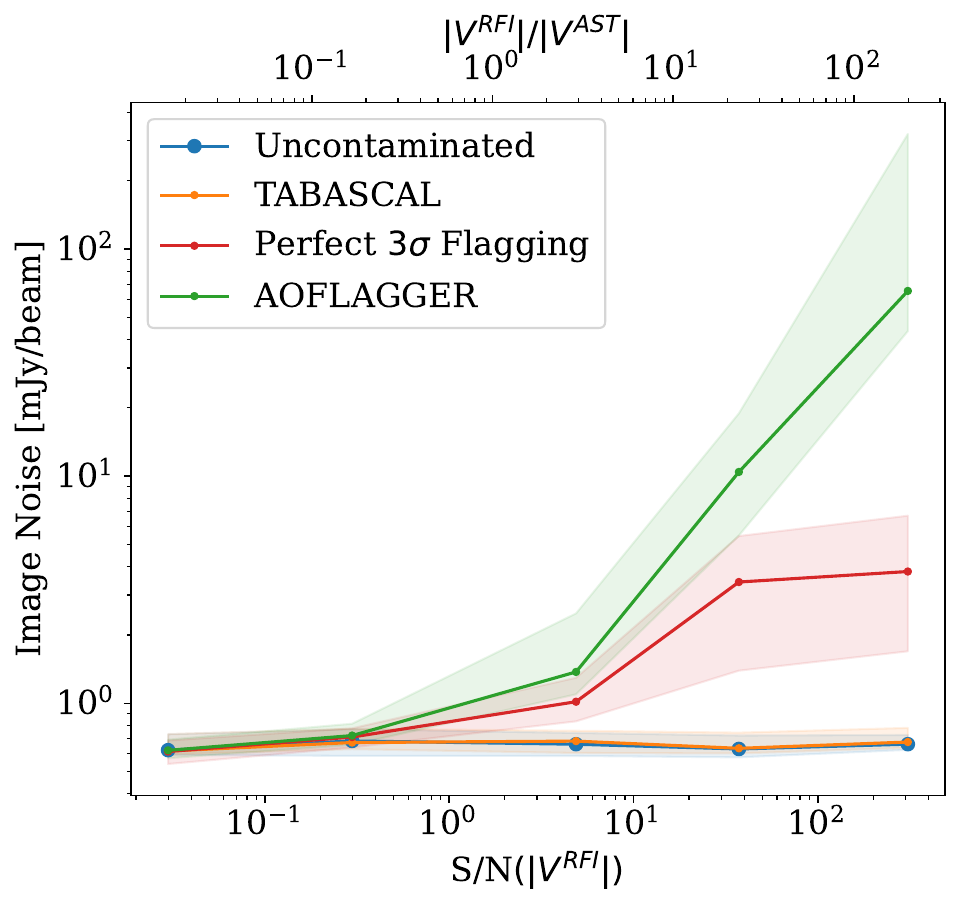}
  \caption{Image noise calculated from the residuals after using WSCLEAN for imaging vs RFI S/N (lower x-axis) and ratio of RFI-to-Astronomical visibility strength (upper x-axis). The solid lines with dots represent the median within an RFI bin containing roughly 16 images each. The shaded region represents the 68\% uncertainty interval over the observations in the bin. TABASCAL (orange) is statistically consistent with the uncontaminated case showing successful signal recovery despite the RFI contamination across the entire range of RFI strengths. In this simulation the astronomical visibilities are chosen to be close in amplitude to the visibility noise so the upper and lower x-axis scales are similar but not identical.}
  \label{fig:image_noise}
\end{figure}

A key metric in evaluating image quality in radio astronomy is the image noise. In this work, the image noise is calculated by taking the standard deviation of the residual image after running WSCLEAN. We found very similar estimates from PYBDSF \citep{mohan2015pybdsf}. Figure \ref{fig:image_noise} shows the image noise for all four cases where observations have been binned according to the mean RFI S/N. The uncontaminated case, of course, does not include any RFI. Therefore, when we show the uncontaminated as a function of the RFI S/N, we are displaying the results from the same observations as the other cases (i.e. we have the same pointing and visibility noise realisation). The dotted lines give the median across the images in a bin and the shaded regions give the 68\% uncertainty interval. We see that TABASCAL (orange) is statistically consistent with the uncontaminated case for all RFI S/N bins. In contrast, the flagging cases are only consistent with the uncontaminated case when the RFI S/N is below 1. Above an S/N of 1, we start to see the flagging cases deviate significantly from the uncontaminated case. The perfect $3\sigma$ flagging case performs significantly better than AOFLAGGER on this particular metric which is more than likely due to AOFLAGGER not flagging all of the RFI in the data as was shown in the middle panel of Fig. \ref{fig:tab_vis_error}. Nonetheless, for RFI S/N values greater than approximately 100, the image noise for perfect $3\sigma$ flagging is at least an order of magnitude greater than for TABASCAL.   

\subsection{Point source recovery}\label{sec:pnt_src}

Although the TABASCAL algorithm we have presented is fully general, to demonstrate its  performance we want a specific test case with clear metrics.  For this purpose we choose point-source imaging which also has many science applications such as the radio luminosity function (RLF) \citet{Sadler2013, Simpson2017, Mauch2020, Tucci2021}.

In this section, we evaluate the performance of our four cases with respect to point source recovery. The images generated for all 79 simulations from the previous section were used here. For this analysis, we consider three metrics to be of primary importance: completeness, purity, and flux estimation error. 

To extract a source catalogue from our images to compare against the input (true) source catalogue we used PYBDSF \citep{mohan2015pybdsf} with the \texttt{thresh\_isl} and \texttt{thresh\_pix} set to 1.5. To ensure suitable statistics, we generated point sources with a minimum source separation of approximately 3 beam widths and fluxes with an expected S/N of greater than 10. Both of these values are estimated based on an uncontaminated observation. However, both the beam width and image noise are expected to be larger in images with significant amounts of flagged data leading to potentially overlapping sources. We used a $5\sigma$ cut (based on the theoretical image noise) on the PYBDSF catalogues to ensure valid source detections. Afterwards, these were matched with the true source catalogues using \texttt{match\_to\_catalog\_sky} from ASTROPY \citep{Astropy2013, Astropy2018, TheAstropyCollaboration2022}. Source pairings were only considered a match if the angular separation was less than the beam width. Finally, any true source that was matched to more than one detected source was cut to only include the closest match, in angular separation, to the true source. 

\begin{figure}
  \centering
  \includegraphics[width=0.45\textwidth]{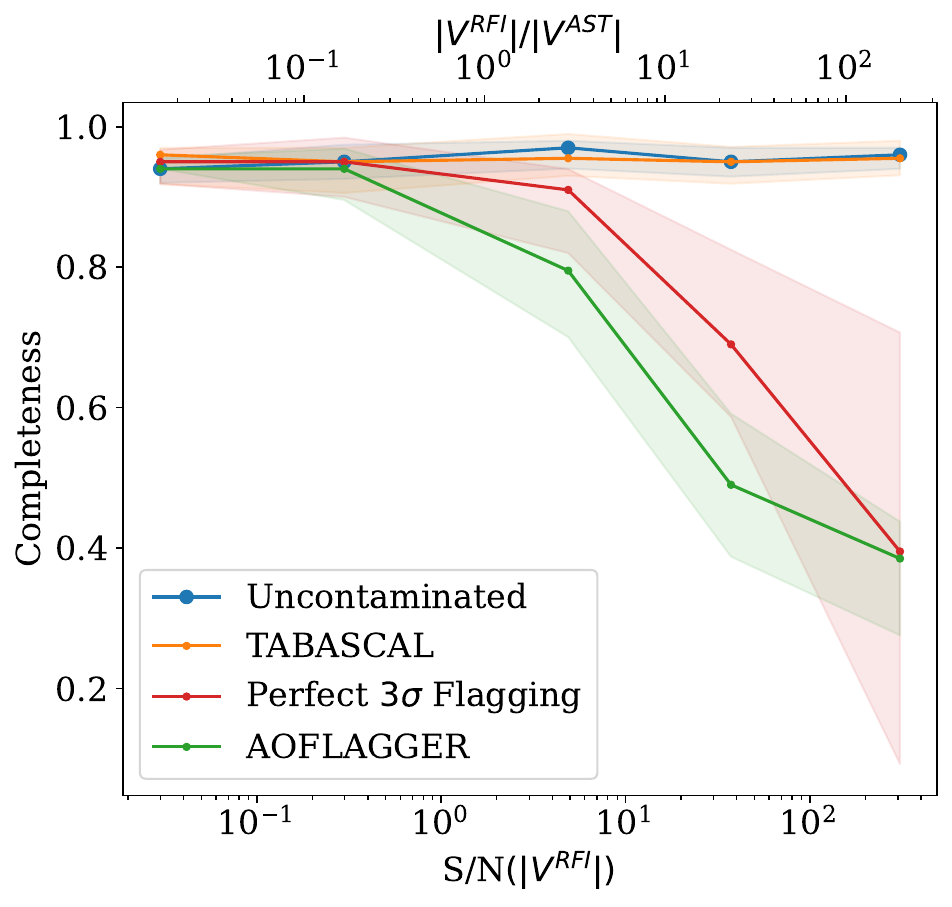}
  \caption{Completeness statistics of source catalogues across RFI S/N bins. Solid lines with dots indicate the median within a bin. The shaded region indicates the 68\% uncertainty interval. Each bin consists of statistics from approximately 16 images for each case.}
  \label{fig:completeness}
\end{figure}

\begin{figure}
  \centering
  \includegraphics[width=0.45\textwidth]{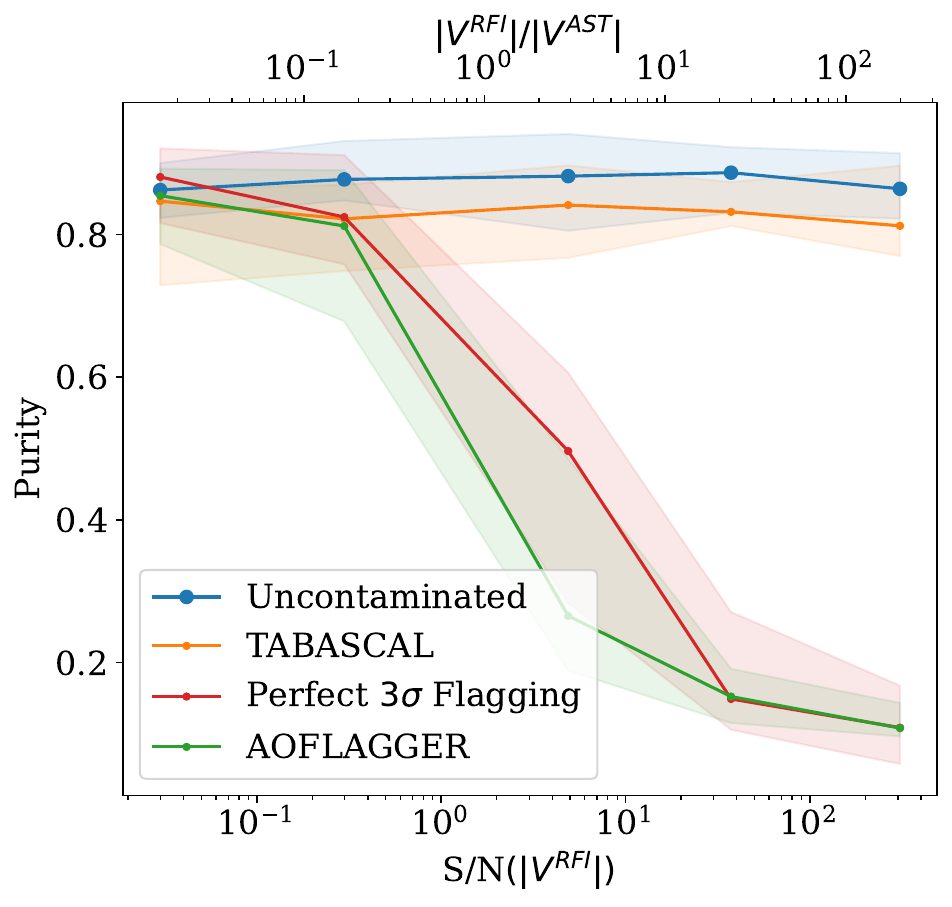}
  \caption{Purity statistics of source catalogues across RFI S/N bins. Solid lines with dots indicate the median within a bin. The shaded region indicates the 68\% uncertainty interval. Each bin consists of statistics from approximately 16 images for each case.}
  \label{fig:purity}
\end{figure}

\begin{figure}
  \centering
  \includegraphics[width=0.45\textwidth]{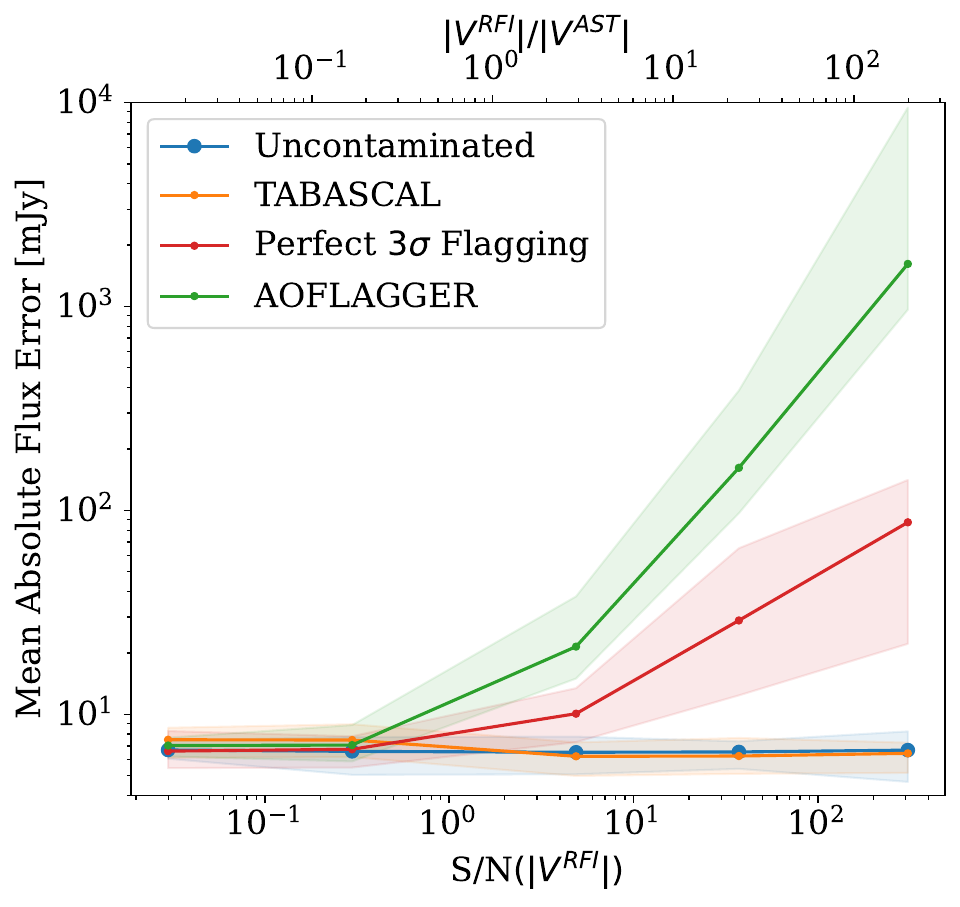}
  \caption{Mean absolute flux error with respect to all matched sources across RFI S/N bins. Solid lines with dots indicate the median within a bin. The shaded region indicates the 68\% uncertainty interval. Each bin consists of statistics from approximately 16 images for each case. TABASCAL (orange) is statistically consistent with the uncontaminated case.}
  \label{fig:flux_error}
\end{figure}

Catalogues for all observations have been binned into five RFI S/N bins. In the following \Cref{fig:completeness,fig:purity,fig:flux_error}, the solid lines with dots represent the median within a bin and the shaded region represents the 68\% uncertainty interval within a bin. Each bin consists of approximately 16 observations. The same imaging and source extraction parameters were used for all datasets across all four cases. 

Completeness, often referred to as recall, is defined as
\begin{equation}
    \text{Completeness} = \frac{\text{TP}}{\text{TP} + \text{FN}},
\end{equation}
where TP stands for true positive (the detected sources that are in the true catalogue) and FN stands for false negative (the undetected sources that are in the true catalogue). Therefore, completeness refers to how complete our detected source catalogue is. Figure \ref{fig:completeness} shows the completeness statistics for our four cases binned with respect to RFI S/N. TABASCAL shows comparable performance to the uncontaminated case for all RFI S/N bins. Both flagging cases have the same performance as TABASCAL for RFI S/N below 1 and start to deviate above this. Significant performance degradation is seen at RFI S/N above 10 with AOFLAGGER consistently performing worse than perfect $3\sigma$ flagging as expected. 

Purity, often referred to as precision, is defined as
\begin{equation}
    \text{Purity} = \frac{\text{TP}}{\text{TP} + \text{FP}},
\end{equation}
where TP is as before and FP stands for false positive (the detected sources that are not part of the true catalogue). Therefore, purity refers to how pure our detected source catalogue is. Figure \ref{fig:purity} shows the purity statistics for our four cases binned with respect to RFI S/N. TABASCAL shows statistically comparable performance to the uncontaminated case across all RFI S/N bins. Again, the flagging cases perform comparably to TABASCAL for RFI S/N below 1. However, on this metric the flagging cases so show very steep performance degradation with increasing RFI S/N. This is likely due to the significant image artefacts that can be seen in Figures \ref{fig:img_comp1} and \ref{fig:img_comp10} for both flagging cases. The images shown are not hand picked and therefore show a representative sample of the broader dataset. At an RFI S/N level of about 30, more than 80\% of the sources detected in the flagging cases are fake sources. These are fake sources that exist at a greater than $5\sigma$ level relative to their image noise, not the theoretical image noise.

We note here that all images appeared to show some low level of ghost sources. This was true for all cases. This is likely due to the particular set of imaging parameters coupled with the sparse $uv$ sampling in these observations. This manifests in slightly lower than perfect purity scores considering the $5\sigma$ flux cuts that were used along with only including $10\sigma_I$ flux sources. In spite of this, we believe these results still give us valuable insight into the performance of TABASCAL and other methods as all cases were affected in the same way. This is further boosted by the fact that the theoretical image noise was attained for all cases with RFI S/N below 1 and for TABASCAL and the uncontaminated across all RFI S/N levels.

Figure \ref{fig:flux_error} shows the mean absolute flux error of the matched sources from the detected catalogue. The solid lines with dots represent the median value within an S/N bin, and the shaded region shows the 68\% uncertainty interval. Each bin consists of results from approximately 16 images for each case considered. The mean absolute flux error (MAFE) is calculated as
\begin{equation}
    \text{MAFE} = \frac{1}{N_s} \sum_{i=1}^{N_s} \lvert \hat{S}_i - S_i \rvert,
\end{equation}
where $\hat{S}_i$ is the predicted total flux for matched source $i$, $S_i$ is the true flux, and $N_s$ is the number of matched sources from a single image. TABASCAL shows comparable MAFE performance to the uncontaminated case across all RFI S/N levels. Much like all other metrics shown, the flagging cases show comparable performance to TABASCAL at S/N levels below 1 and then quickly deteriorate for RFI S/N levels above this. As with all the other metrics shown AOFLAGGER performs worse than the perfect $3\sigma$ flagging case as expected. On the MAFE metric, AOFLAGGER shows a faster degradation in performance, with increasing RFI S/N, compared to the other metrics relative to TABASCAL. This is likely due to the image artefacts that arise. Examples of these image artefacts are shown in Figures \ref{fig:img_comp1} and \ref{fig:img_comp10}. When considering the purity performance, shown in Fig. \ref{fig:purity}, in conjunction with the MAFE results this conclusion seems likely. At RFI S/N levels of 40, perfect $3\sigma$ flagging performs around five times worse than TABASCAL with AOFLAGGER performing roughly 30 times worse.

\subsection{Gain phase calibration}\label{sec:phase_cal}

\begin{figure}
  \centering
  \includegraphics[width=0.5\textwidth]{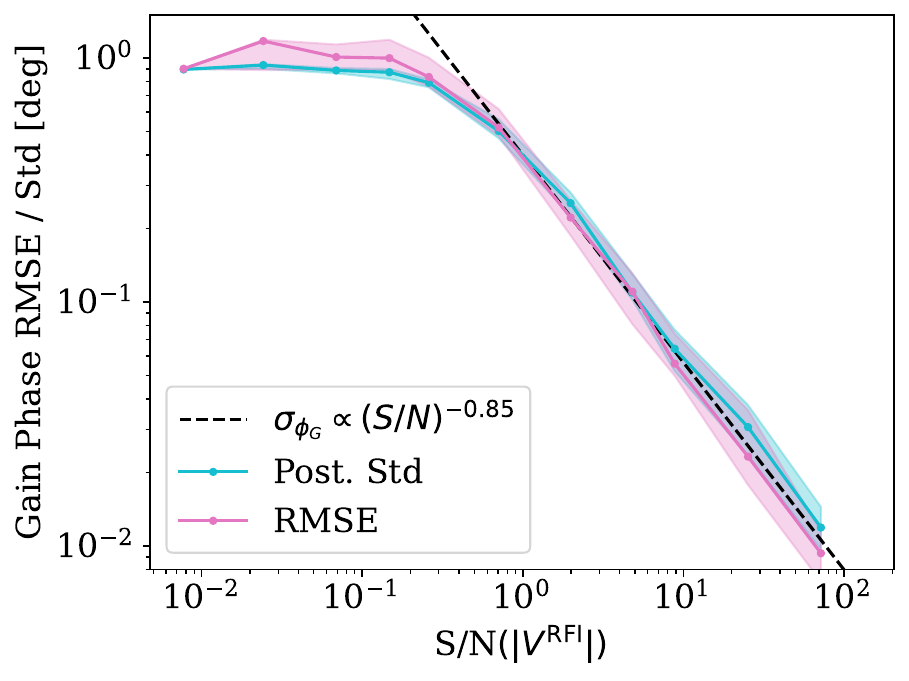}
  \caption{Evidence that TABASCAL achieves phase calibration off the strong RFI sources: A comparison of the root mean squared error (RMSE) in gain phase estimates and the posterior standard deviations from TABASCAL. TABASCAL shows statistically rigorous phase calibration capabilities that depend on the RFI S/N. Above an S/N of 1, phase constraints are inversely proportional to the RFI S/N. Below an S/N of 1, phase solutions are found within the prior information given.}
  \label{fig:g_phase}
\end{figure}

Figure \ref{fig:g_phase} shows TABASCAL's phase calibration capabilities are shown with increasing RFI S/N. The root mean squared error (RMSE) in the gain phase, for a single observation, is calculated as follows:
\begin{equation}
    \text{RMSE} = \sqrt{\frac{1}{N_A N_T} \sum_{p,i} \left( \hat{\phi}_{G_p}(t_i) - \phi_{G_p}(t_i) \right)^2 },
\end{equation}
where $\hat{\phi}_{G_p}(t_i)$ is the TABASCAL estimate of the gain phase on antenna $p$ and time $t_i$, and $\phi_{G_p}(t_i)$ is the true gain phase value. The posterior standard deviations for a single observation are also combined in a root mean squared sense. Results from all observations are binned according to the mean RFI S/N values in each. The median and 68\% uncertainty interval, across observations, are represented with a solid dotted line and shaded region respectively. The RMSE shows good agreement with the posterior errors indicating that TABASCAL's MAP estimates and errors are statistically consistent. The proportional decrease between the posterior uncertainties and the RFI S/N indicates that the phase calibration capabilities of TABASCAL are a legitimate feature of the model. The prior distribution on the gains was chosen to have standard deviations of 1\% and 1 degree in the amplitude and phase respectively. The mean of the prior distribution is set by taking a sample from a distribution with equivalent standard deviation centred on the true value. Therefore, the prior distribution is not centred on the true value but is statistically consistent with the true value.

\section{Discussion}\label{sec:discussion}

\subsection{Considerations for real data}

In this work, we  tested TABASCAL on simulations that include the dominant factors required to mimic real data. However, there are some scenarios and effects that have not been included that may be present in real data that need to be considered. The main considerations are: (1) multipath propagation effects, (2) non-linear telescope response, and (3) unmodelled sources of RFI. We discuss these below.

In our simulations, we only considered the direct line of sight between the RFI source and the antennas. Although our model for the RFI signal for each source at each antenna allows for some path difference, it is not flexible enough to account for multipath effects such as diffractions and reflections from environmental features. The contribution of these effects is expected to be negligible to non-existent for satellite-based RFI sources. Although TABASCAL is general and can be applied to nearly all off-site sources of RFI, satellite-based sources are the main focus of this work and of the largest growing concern given the advent of satellite mega-constellations \citep{Witze2025}.

Our simulations (and the forward model) assume the telescope's response to the RFI signal to be linear. This is known to be an invalid assumption when the RFI signal is strong enough. This is especially true when the RFI signal is so strong that the receiver is saturated and any signal recovery is impossible. In our Bayesian forward model, the potential for non-linear effects is not taken into consideration. In principle, this can be modelled and included into the simulations and forward model. Currently, however, this stands as a limitation for TABASCAL, although there is interest in researching this direction. For now, there is still a large range of RFI signal amplitudes, where the receiver remains linear, that TABASCAL is applicable to and will greatly benefit the sensitivity of observations.

A valid concern with a model-based approach like TABASCAL is when unmodelled sources of RFI, such as other satellites, planes, broadcast towers, and so on are present in the data. An impact analysis of such a case has not been performed in this work. Given that TABASCAL is a model-based approach, any unmodelled signal from other sources will lead to a degradation of the results. This will be apparent from the increased $\chi^2$ per data point when the signal contribution is comparable to or greater than the data noise. A residual fast varying, `model-free' signal component could be included in the future, however, this is likely to degrade the astronomical signal recovery. The best solution is to determine the sources of unmodelled signals and include them into TABASCAL's model.

\subsection{Effect of the prior and reduced information}

To establish the limits of the model used within TABASCAL a number of tests were run with respect to reducing prior information. In this section, we also discuss the limitations of the priors and what assumptions are present.

\subsubsection{Astronomical visibility prior}\label{sec:ast_prior}

Variation of the astronomical visibility prior can be done through the three parameters defined in Equation \eqref{eq:power_spec}, namely: $P_0$, $\eta_0$, and $\gamma$. First, $P_0$ defines the prior variance of the astronomical visibility signal. The only significant effect found on the solutions when varying $P_0$ was when it was made too small. This effectively excludes higher signal amplitudes leading to an under-estimation of the visibilities. Then, $\eta_0$ can be calculated based on the geometry of the telescope, as per Equation \eqref{eq:ast_fringe}. The main assumption in this is that no strong sources are present in the sidelobes. In the presence of such sources, the same equation can be used but the angular separation between the furthest off-axis source and pointing direction should be used instead of the telescope FoV radius. Alternatively, such sources could be modelled separately within the TABASCAL framework. The traditional method for dealing with this is source peeling. We did not test this scenario. For mid-frequency observations where there is a limited FoV, adapting Equation \eqref{eq:ast_fringe} is expected to work well. For wide FoV observations the increased $\eta$, namely, widening of the fringe filter, could lead to RFI signal leakage into the predicted astronomical signal.

As stated previously, $\gamma$ controls the smoothness of the solutions. During our investigations, we found that values of $\gamma \leq 3$ led to a bias in the gain amplitudes at RFI S/N below 1. The gain amplitudes would consistently be overestimated in this range, which led to an underestimation of the astronomical visibilities and subsequently the source flux recovery. For RFI S/N values above 1, this bias went away, indicating that RFI signal plays a role in constraining gain amplitudes to within the prior distribution. $\gamma$ plays the role of the steepness on the edges of the effective fringe filter induced by the astronomical visibility prior. A steeper drop-off (larger $\gamma$) led to more consistent and less biased results. We note that we did not perform any tests with $\gamma>5$. However, a previous (less efficient) implementation where an SE kernel in the time domain was used, did not exhibit this bias. The SE kernel is the limiting case of our power spectrum model where $\gamma\rightarrow\infty$. $\gamma$ is required in TABASCAL's current implementation for numerical stability. However, this will be improved in the future. Our recommendation is to use as large a $\gamma$ as possible.   

Tests of limiting cases were performed with respect to the presence or absence of astronomical and RFI signals. We found that excluding astronomical signal in the data led to the same results as have been presented in this paper. This was the case both when including or excluding RFI parameters. Excluding RFI signal led to equivalent results to those of the RFI S/N $<10^{-1}$. The same results were found both when including or excluding the astronomical visibility parameters. 

\subsubsection{RFI signal prior and old TLEs}\label{sec:rfi_prior}

The prior on the RFI signal consists of two parameters, namely, $\sigma^2_{A_\text{RFI}}$ and $\tau_\text{RFI}$. $\sigma^2_{A_\text{RFI}}$ defines the prior variance of the RFI signal at an antenna. This roughly corresponds to the standard deviation in RFI visibility amplitude for a single source. Throughout this work, we have used $\sigma^2_{A_\text{RFI}} = 10^4$ Jy, which is larger than any RFI signal used in this work. Reducing this below the RFI strength in the data leads to an underestimation of the signal and, subsequently, a $\chi^2$ per data point above 1.1 \footnote{$\chi^2$ per data point below 1.1 is the metric used for convergence and a successful fit.}. $\tau_\text{RFI}$ defines the correlation time of the RFI signal. Alternatively, $1/2\pi \tau_\text{RFI}$ can be thought of as the width of a fringe filter (relative to the array centre) about the estimated direction of the RFI source. This allows for the RFI signal parameter $A_\text{RFI}$ to fit for a range of residual signal effects as described in Section \ref{sec:rfi_signal}. Equation \eqref{eq:rfi_l} shows the calculation of $\tau_\text{RFI}$ to account for, at minimum, the primary beam modulation. For our simulations this was calculated to be $\tau_\text{RFI} = 24$ s corresponding to $1/2\pi \tau_\text{RFI} = 6.6$ mHz or 13 mHz on a baseline. When testing TABASCAL with one day old TLEs, corresponding to an average position error of 190 m, we found that  $1/2\pi \tau_\text{RFI} = 6.6$ mHz was enough to account for the position errors in 71\% of observations compared to 86\% in the base case. We did not test the robustness of solutions to ionospheric fluctuations or beam irregularities. This is left for future work. When reducing $\tau_\text{RFI}$ to 5 seconds (32 mHz) we found statistically comparable results on all metrics, across all RFI S/N levels relative to the base case. We did not test the lowering of $\tau_\text{RFI}$ in conjunction with old TLE data to see if the convergence rate was increased.

\subsubsection{Gain prior}\label{sec:gain_prior}

The prior on the gains consists of two parameters each for the amplitudes and phases, the standard deviation ($\sigma_{|G|},\sigma_{\phi_G}$) and correlation time ($\tau_{|G|},\tau_{\phi_G}$). When increasing $\sigma_{|G|}$ from 1\% to 5\%, we found image noise and completeness to be statistically comparable to the base case. We found flux recovery and purity to be slightly diminished across the RFI S/N range, as expected given the apparent lack of amplitude calibration capabilities. When increasing $\sigma_{\phi_G}$ from 1$^\circ$ to 5$^\circ$, we found decreased performance in the astronomical source flux recovery at RFI S/N below 1 and equal performance to the base case above 1 thanks to the phase calibration capabilities leveraging the RFI signal. The recovered astronomical source fluxes were underestimated (below RFI S/N 1) likely due to the signal decoherence smearing the astronomical sources in the image. When decreasing $l$ to 30 seconds compared to 3 hours, for both the amplitudes and phases, we found no effect on the results. This is an indication that TABASCAL would be able to handle more rapid gain fluctuations; however, this was not tested. 

\subsection{Computational costs and potential improvements}

The computational costs of this method are dominated by the calculation of the RFI visibilities from the interpolated RFI signal, and phases, at each antenna. The number of floating-point operations (FLOPS) can be calculated as 
\begin{equation}
    N_\text{FLOPS} \approx 2 N_A^2 N_f N_\text{RFI} \Delta T \nu_s,
\end{equation}
where $\Delta T$ is the total observation time considered and $\nu_s$ is the sampling frequency determined by the strongest RFI source. The memory requirements, at single precision, are approximately $4N_\text{FLOPS}$ Bytes. The optimisation time, namely, the cost of a MAP estimate, assuming 2000 gradient steps, is approximately a microsecond per FLOP. Here we are using FLOPS to define the problem size and not the actual number of FLOPS performed by the processor. An accurate TLE estimate and good initialisation of the RFI signal parameters requires only 1000-2000 gradient steps. An outdated TLE estimate from the day before typically requires 6000 gradient steps.

In our work we have used a single Nvidia A100 GPU. The problem size used in this work consists of $N_A=32$, $N_f=1$, $N_\text{RFI}=$ 2 - 9, and $\nu_s \Delta T = $ 1300 - 6200. This resulted in runtimes between 14 and 140 seconds for a 15 minute observation, namely, 6x - 60x faster than the observation time. Extrapolating this to an SKA Mid size telescope of $N_A=192$, an average of $N_\text{RFI}=10$ at mid-earth orbit, and mean RFI S/N of 100 ($\nu_s=10$ Hz), we expect approximately 7 seconds of compute per second of observation for each contaminated channel.

The above calculation has ignored the increased baseline length present in a larger array. Longer baselines lead to an increased fringe frequency as shown in Equation \eqref{eq:sat_fringe}. Increased fringe rates also lead to more signal decoherence, so these baselines would not be as greatly affected by RFI. Baselines on which the RFI fringe rate is so large that the RFI signal is below the noise already could be excluded. Traditional fringe filtering methods such as that proposed by \citet{Offringa2012} will be very effective on these longer baselines. TABASCAL best serves the shorter baselines where fringe rates are lower, making it harder for traditional methods to separate the astronomical and RFI signals.

The computational efficiency of TABASCAL can be improved in a number of ways, paving the way for its widespread use. Currently, the sampling frequency $\nu_s$ is chosen based on the fringe rate, $\nu_f$, of the fastest fringing source, additionally, the expensive portion (complex multiplication between antenna signals) does not take into account the fringe rate on a particular baseline. Ideally, the sampling frequency for each RFI source and baseline would be used to calculate their associated visibility value, after which, per source visibilities can be summed to give the total RFI visibility. This is the single largest computational improvement that can be made and would open up TABASCAL to be used across an entire array with very long baselines efficiently. The next largest computational improvement would be in the interpolation of the RFI signal. This can be done through the use of Fourier space GPs, as is used for the astronomical visibilities. Currently, interpolation is performed in the time domain where a full matrix multiplication is required at a cost of $\mathcal{O} \left( \nu_s\Delta T^2/\tau_\text{RFI} \right)$. A Fourier domain interpolation would only cost $\mathcal{O} \left( N\log(N)  \right)$ with $N=\nu_s\Delta T$. This is only beneficial when $\Delta T/\tau_\text{RFI} > \log(\nu_s\Delta T)$ which is almost always the case. 

\subsection{Improved prior information}

Increased prior knowledge can improve the computational efficiency of TABASCAL, the stability of the algorithm, as well as potentially its capabilities for amplitude calibration. As stated in the previous section, better TLE estimates lead to faster convergence. This is a statement about the accuracy in positional information of the RFI sources. In some sense, this can be thought of as one component of phase calibration in the direction of the RFI source. If the primary beam is well modelled and included as an independent factor in TABASCAL's model, $\tau_\text{RFI}$ can be increased as the RFI signal variation due to the source's movement through the sidelobes does not need to be accounted for in the GP signal model. Increasing our prior knowledge about the intrinsic RFI signal of a given source could even lead to amplitude calibration capabilities using the RFI signal. For example, knowledge of the beam pattern and pointing direction of an RFI signal can help us further constrain $\tau_\text{RFI}$. Improvements like these can help TABASCAL to narrow the effective phased-up fringe filter and improve its convergence and success rate beyond what has been shown in this work.

Currently TABASCAL uses an exceptionally wide prior on the astronomical visibilities about 0. In the future, an informative prior could be used based on imaging of a neighbouring uncontaminated channel. This could further improve the convergence and success rate of TABASCAL.

\section{Conclusions}\label{sec:conclusions}

\subsection{Method summary}

TABASCAL solves the problem of estimating phase calibrated astronomical visibilities in the presence of RFI sources that follow predictable trajectories. To achieve this, TABASCAL requires knowledge of the RFI sources present in the data as well as a reasonable prior estimate of the RFI trajectories. Furthermore, the priors used on the RFI and astronomical signals in TABASCAL can be thought of as independent statistical fringe filters, each phased-up into the direction of the corresponding sources. This gives TABASCAL the power to effectively separate the RFI and astronomical signals while accounting for differential fringe rates in all considered directions. 

\subsection{Key results summary}

Overall summary: TABASCAL is able to `see through' predictable RFI (satellites, ground stations, etc.) as if it were not there. This may be a practical solution to the challenge of ever more satellites. The key results are summarised below. 
\paragraph{Astronomical visibility recovery:}
\begin{itemize}
    \item Accuracy: TABASCAL achieves accurate recovery of astronomical visibilities, even under strong RFI contamination. Predictions are highly Gaussian out to $3\sigma$. 
    \item Comparison to flagging: Both AOFLAGGER and perfect $3\sigma$ flagging struggle at RFI S/N strengths above 1, failing to match TABASCAL's accuracy due to incomplete RFI mitigation.
\end{itemize}

\paragraph{Imaging performance:}
\begin{itemize}
    \item Image quality: TABASCAL produces images with noise and artefacts comparable to the uncontaminated case across all RFI levels. In contrast, perfect $3\sigma$ flagging and AOFLAGGER images exhibit significantly higher noise and artefacts, especially at high RFI S/N.
    \item Noise metrics: Image noise for TABASCAL remains statistically consistent with the uncontaminated case, unlike flagging approaches, which show significant noise increases at RFI S/N greater than 1.
\end{itemize}

\paragraph{Point source recovery:}
\begin{itemize}
    \item Completeness: TABASCAL maintains high completeness across all RFI levels, comparable to the uncontaminated case and significantly better than flagging methods.
    \item Purity: TABASCAL consistently outperforms flagging methods with performance consistent with uncontaminated.
    \item Flux estimation: Flux estimation errors are low for TABASCAL, with performance comparable to the uncontaminated benchmark.
\end{itemize}

\paragraph{Calibration:}
\begin{itemize}
    \item Amplitude: TABASCAL includes gain amplitude parameters, however, the errors in the estimated gain amplitudes remain at the same level as is provided in the prior distribution. Therefore, TABASCAL shows no amplitude calibration capability in the absence of no known calibrator sources. The direct inclusion of an astronomical source model is expected to lead to amplitude calibration capabilities, namely, the amplitude estimates would have lower error than given in the prior distribution. 
    \item Phase: TABASCAL is able to leverage the RFI signal to constrain phase calibration solutions in a statistically consistent manner. Phase calibration constraints are shown to be directly proportional to the RFI S/N. 
\end{itemize}

\paragraph{Stress testing:}
\begin{itemize}
    \item Robustness: TABASCAL shows resilience under stress tests involving old TLE data and variations in prior hyperparameters, maintaining effective RFI mitigation and astronomical visibility recovery.
\end{itemize}

Overall, TABASCAL demonstrates remarkable, robust performance, achieving a level of visibility and image quality close to the uncontaminated ideal and outperforming traditional flagging methods in all scenarios for moderate and strong RFI contamination (S/N $\geq$ 1). For low levels of RFI contamination, all methods demonstrate an approximately equal level of performance. In future works, TABASCAL will be tested on real astronomical data and further computational improvements will be made. We wish to see TABASCAL become part of the standard data reduction pipeline in the future, helping to tackle the ever-growing impact of RFI, especially that of satellite mega-constellations.

\begin{acknowledgements}
    We thank members of the SARAO Data Science team, Radio Astronomy Research Group at SARAO, Niruj Mohan and Robert Mueller for useful discussions. CF and MK acknowledge funding by the Swiss National Science Foundation. We also acknowledge the support of the South African Radio Astronomy Observatory. This research has been conducted using resources provided by the Science and Technology Facilities Council (STFC) through the Newton Fund and SARAO.
\end{acknowledgements}

\section*{Data availability}

The simulation and visibility recovery codes are available on their respective GitHub repositories \href{https://github.com/chrisfinlay/tab-sim}{tab-sim} \footnote{\href{https://github.com/chrisfinlay/tab-sim}{https://github.com/chrisfinlay/tab-sim}} and \href{https://github.com/epfl-radio-astro/tabascal}{tabascal} \footnote{\href{https://github.com/epfl-radio-astro/tabascal}{https://github.com/epfl-radio-astro/tabascal}}.

\bibliographystyle{aa} 
\bibliography{main} 

\begin{appendix}

\FloatBarrier
\section{Covariance estimation}\label{app:cov_est}

We derive the method for scalable covariance estimation here. We first start by linearising our forward model of the visibilities $\bm{V}$ as a function of the parameters $\bm{\theta}$ about the MAP point $\hat{\bm{\theta}}$.
\begin{align}
    \bm{V} \left( \bm{\theta} \right) &= \bm{V} \left( \hat{\bm{\theta}} + \delta \bm{\theta} \right) \\
              &= \bm{V} \left( \hat{\bm{\theta}} \right) + \bm{J} \delta\bm{\theta} + \mathcal{O} \left( \delta\bm{\theta}^2 \right),
\end{align}
and therefore by defining $\delta \bm{V} = \bm{V} \left( \bm{\theta} \right) - \bm{V} ( \hat{\bm{\theta}} ) + \bm{J}\hat{\bm{\theta}}$, we get the linearised model as
\begin{equation}
    \delta \bm{V} = \bm{J} \bm{\theta},
\end{equation}
where $\bm{J}=\frac{\partial \bm{V}}{\partial\bm{\theta}}$ is the Jacobian. The prior distribution on $\bm{\theta}$ is 
\begin{equation}
    p(\bm{\theta}) = \mathcal{N} \left( \bm{\mu}_\Pi, \bm{\Sigma}_\Pi \right),
\end{equation}
where $\bm{\mu}_\Pi$ and $\bm{\Sigma}_\Pi$ are the prior mean and covariance, respectively. The likelihood, assuming Gaussian noise with covariance $\bm{\Sigma}_N$, is then
\begin{equation}
    p(\delta \bm{V} | \bm{\theta}) = \mathcal{N} \left( \bm{J}\bm{\theta}, \bm{\Sigma}_N \right).
\end{equation}
Given that we are working with a linearised model, the posterior distribution is therefore a Gaussian distribution and is defined as follows
\begin{equation}
    p \left( \bm{\theta} | \delta \bm{V} \right) = \mathcal{N} \left( \hat{\bm{\theta}}, \hat{\bm{\Sigma}} \right),
\end{equation}
where 
\begin{equation}\label{eq:post_cov}
    \hat{\bm{\Sigma}} = \left( \bm{J}^T \bm{\Sigma}_N^{-1} \bm{J} + \bm{\Sigma}_\Pi^{-1} \right)^{-1}
\end{equation}
and
\begin{equation}\label{eq:post_mean}
    \hat{\bm{\theta}} = \hat{\bm{\Sigma}} \left( \bm{J}^T \bm{\Sigma}_N^{-1} \delta \bm{V} + \bm{\Sigma}_\Pi^{-1} \bm{\mu}_\Pi  \right).
\end{equation}
We never actually evaluate Equation \eqref{eq:post_mean} as we already have this point, the MAP $\hat{\bm{\theta}}$ from our optimisation. However, if we add Gaussian perturbations $\Delta\bm{\psi} \sim \mathcal{N} \left( \bm{0}, \bm{\Sigma}_N \right)$ and $\Delta\bm{\phi} \sim \mathcal{N} \left( \bm{0}, \bm{\Sigma}_\Pi \right)$ to $\delta \bm{V}$ and $\bm{\mu}_\Pi$ respectively, we obtain the following,
\begin{align}\label{eq:cov_sampling}
    \hat{\bm{\theta}} + \Delta\bm{\theta} &= \hat{\bm{\Sigma}} \left( \bm{J}^T \bm{\Sigma}_N^{-1} \left ( \delta\bm{V} + \Delta\bm{\psi} \right) + \bm{\Sigma}_\Pi^{-1} \left(\bm{\mu}_\Pi + \Delta\bm{\phi} \right)  \right), \\
    \implies \Delta\bm{\theta} &= \hat{\bm{\Sigma}} \left( \bm{J}^T \bm{\Sigma}_N^{-1} \Delta\bm{\psi} + \bm{\Sigma}_\Pi^{-1} \Delta\bm{\phi} \right).
\end{align}
It can be easily shown that $\mathbb{E} \left[ \hat{\bm{\theta}} + \Delta\bm{\theta} \right] = \hat{\bm{\theta}}$ and $\mathbb{E} \left[ \Delta\bm{\theta} \Delta\bm{\theta}^T \right] = \hat{\bm{\Sigma}}$ as desired. Therefore, we now have a way to draw samples from our approximated posterior with which we can estimate covariances and marginal uncertainties easily. This process of drawing samples by Gaussian perturbations is taken from \citet{papandreou2010gaussian}. 

Given that we have standardised our parameter space and the measurements have independent noise, as is done in \citet{knollmuller2019metric}, we therefore have 
\begin{align}
    \bm{\Sigma}_\Pi &= \bm1, \\
    \bm{\Sigma}_N &= \sigma_n^2\bm{1}.
\end{align}
We can see that obtaining the samples from $\bm{\Sigma}_N^{-1} \Delta\bm{\psi}$ and $\bm{\Sigma}_\Pi^{-1} \Delta\Phi$ is both easy and scales linearly with the number of parameters $N_P$. The expensive part in terms of memory and computation comes from applying the Jacobian $\bm{J}$ and the inverse posterior information $\hat{\bm{\Sigma}}$ at the MAP location. These are a Jacobian-vector product (JVP) and matrix-vector product (MVP) respectively and can be defined implicitly, namely, without the need to evaluate the Jacobian of size $N_P \times N_D$ and the posterior information of size $N_p \times N_P$ explicitly. Additionally, evaluating Equation \eqref{eq:cov_sampling} requires the inversion of the posterior information matrix, $\hat{\bm{\Sigma}}^{-1}$, which would require $\mathcal{O}(N_P^3)$ computation typically. However, since it is by definition a symmetric operator, we can use the conjugate gradient method \citep{shewchuk1994introduction} to apply its inverse to a vector. The benefit of conjugate gradient is we do not need to know the posterior information explicitly, and the inversion can be done in $\mathcal{O}(N_P^2)$ computation or less if an approximate solution is acceptable. Since the model is implemented in JAX \citep{jax2018github}, where the Jacobian-vector product is an integral part of the framework and implicit operators are easily defined, the implementation of the above sampling technique is therefore trivial. The error in the covariance estimate scales inversely to the number of samples taken, as expected. This type of implementation allows us to balance the required accuracy in the covariance estimate with the available computation while remaining scalable to millions of parameters.

\FloatBarrier
\section{Simulation and prior parameters}\label{sec:parameters}

\begin{table}
\caption{Summary of telescope simulation parameters.}
\centering
\def\arraystretch{1.5}
    \begin{tabular}{c|c}
        \multicolumn{2}{c}{\bfseries Telescope} \\
        \hline
        Parameter & Value/Range \\
        \hline\hline
        Latitude & -30$^\circ$ \\
        \hline
        Longitude & 21$^\circ$ \\
        \hline
        \# Antennas ($N_A$) & 32 \\
        \hline
        Dish diameter ($D$) & 13.5 m \\
        \hline
        Frequency ($\nu$) & 1.227 GHz \\
        \hline
        Channel width ($\Delta \nu$) & 209 kHz \\
        \hline
        \# Channels ($N_\nu$) & 1 \\
        \hline
        Visibility noise ($\sigma_n$) & 0.65 Jy \\
        \hline
        Integration time ($\Delta t$) & 2 s \\
        \hline
        Observation time ($\Delta T$)& 15 min \\
        \hline
        \# Time steps ($N_T$) & 450 \\
        \hline
        Sampling frequency ($\nu_s$) & 513 Hz \\
        \hline
        Gain amplitude drift $( \dot{|G|} )$ & 2.4\%.hr$^{-1}$ \\
        \hline
        Gain phase drift $\left( \dot{\phi_G} \right)$ & 1.4$^\circ$.hr$^{-1}$ \\
\end{tabular}
\label{tab:tel_sim}
\end{table}

\begin{table}
\caption{Summary of sky simulation parameters.}
\centering
\def\arraystretch{1.5}
        \begin{tabular}{c|c}
        \multicolumn{2}{c}{\bfseries Sky} \\
        \hline
        Parameter & Value/Range \\
        \hline\hline
        \# Sources & 100 \\
        \hline
        Source type & Point \\
        \hline
        Source flux & 14 mJy - 1 Jy \\
        \hline
        Flux distribution & $\propto S^{-1.6}$ \\
        \hline
        Right Ascension & 0$^\circ$ - 360$^\circ$ \\
        \hline
        Declination & -60$^\circ$ - 30$^\circ$ \\
        \hline
        Local hour angle & (0.5$^\circ$ - 4.3$^\circ$) \\
        \hline
        \# Satellites &  2 - 9 \\
        \hline
        Satellite type & NAVSTAR (GPS) \\
        \hline
        Mean RFI (satellite) flux & $4.2$ mJy - $1.5$ kJy \\
\end{tabular}
\label{tab:sky_sim}
\end{table}

\begin{table}
\caption{Summary of prior parameters used in the results of this work.}
\centering
\def\arraystretch{1.8}
        \begin{tabular}{c|c|c}
        \multicolumn{3}{c}{\bfseries Prior} \\
        \hline
        Parameter Description & Symbol & Value \\
        \hline
        \hline
        \multicolumn{3}{l}{\bfseries Astronomical Visibility} \\
        \hline
        Fringe space variance & $P_0$ & $10^7$ Jy$^2$ \\
        \hline
        Characteristic fringe rate & $\eta_0$ & 1 mHz \\
        \hline
        Smoothness & $\gamma$ & 5 \\
        \multicolumn{3}{l}{\bfseries Gains} \\
        \hline
        Amplitude standard deviation & $\sigma_{|G|}$ & 1\% \\
        \hline
        Amplitude correlation time & $\tau_{|G|}$ & 3 hr \\
        \hline
        Phase  standard deviation & $\sigma_{\phi_G}$ & 1$^\circ$ \\
        \hline
        Phase correlation time & $\tau_{\phi_G}$ & 3 hr \\
        \multicolumn{3}{l}{\bfseries RFI Signal} \\
        \hline
        Signal variance & $\sigma^2_{A_\text{RFI}}$ &  $10^4$ Jy \\
        \hline
        Correlation time & $\tau_{A_\text{RFI}}$ & 24 s \\
\end{tabular}
\label{tab:prior_params}
\end{table}

\end{appendix}

\end{document}